# An Algorithmic Information Calculus for Causal Discovery and Reprogramming Systems

**Hector Zenil^\*[a,b,c,d,e], Narsis A. Kiani\*[a,b,d,e], Francesco Marabita[b,d], Yue Deng[b], Szabolcs Elias[b,d], Angelika Schmidt[b,d], Gordon Ball[b,d], & Jesper Tegnér^[b,d,f]**

a) Algorithmic Dynamics Lab, Center for Molecular Medicine, Karolinska Institutet, Stockholm, 171 76, Sweden
b) Unit of Computational Medicine, Center for Molecular Medicine, Department of Medicine, Solna, Karolinska Institutet, Stockholm, 171 76, Sweden
c) Department of Computer Science, University of Oxford, Oxford, OX1 3QD, UK.
d) Science for Life Laboratory, Solna, 171 65, Sweden
e) Algorithmic Nature Group, LABORES for the Natural and Digital Sciences, Paris, 75006, France.
f) Biological and Environmental Sciences and Engineering Division, Computer, Electrical and Mathematical Sciences and Engineering Division, King Abdullah University of Science and Technology (KAUST), Thuwal 23955–6900, Kingdom of Saudi Arabia

* Shared-first authors
^ Corresponding authors

The Online Algorithmic Complexity Calculator implements the perturbation analysis method introduced in this paper: http://complexitycalculator.com/ and an online animated video explains some of the basic concepts and motivations to a general audience: https://youtu.be/ufzq2p5tVLI

**Abstract:**

**We demonstrate that the algorithmic information content of a system is deeply connected to its potential dynamics, thus affording an avenue for moving systems in the information-theoretic space and controlling them in the phase space. To this end we performed experiments and validated the results on (1) a very large set of small graphs, (2) a number of larger networks with different topologies, and (3) biological networks from a widely studied and validated genetic network (e.coli) as well as on a significant number of differentiating (Th17) and differentiated human cells from high quality databases (Harvard's CellNet) with results conforming to experimentally validated biological data. Based on these results we introduce a conceptual framework, a model-based interventional calculus and a reprogrammability measure with which to steer, manipulate, and reconstruct the dynamics of non-linear dynamical systems from partial and disordered observations. The method consists in finding and applying a series of controlled interventions to a dynamical system to estimate how its algorithmic information content is affected when every one of its elements are perturbed. The approach represents an alternative to numerical simulation and statistical approaches for inferring causal mechanistic/generative models and finding first principles. We demonstrate the framework's capabilities by reconstructing the phase space of some discrete dynamical systems (cellular automata) as case study and reconstructing their generating rules. We thus advance tools for reprogramming artificial and living systems without full knowledge or access to the system's actual kinetic equations or probability distributions yielding a suite of universal and parameter-free algorithms of wide applicability ranging from causation, dimension reduction, feature selection and model generation.**

# 1. Introduction and Preliminaries

How to make optimal predictions about the behaviour of dynamical complex systems is a fundamental problem in science. It remains a challenge to understand and ultimately reprogram the behaviour of such systems with access to only partial knowledge and incomplete or noisy data. Based on established knowledge drawn from the mathematical theories of computability and algorithmic probability that describe the limits of optimal characterization and algorithmic inference, we introduce a conceptual framework, with specific methods and applications that demonstrate the use and advantage of a powerful calculus based on the change of a system's algorithmic content over time and when subject to perturbations.

## 1.1 Causality, Probability and Algorithmic Complexity

The theory of Algorithmic Information[2] provides a definition of what constitutes *a cause* in the real of discrete dynamical systems. Formally, the algorithmic complexity[3,4] of a string s is given by $C(s|e) := \min\{|p| : U(p,e) = s\}$, where p is the program that produces s and halts, running on a (prefix-free[5]) universal Turing machine U with input e which can be empty and is represented simply by $C(s)$. $C(s)$ is the length of the description of the generating mechanism. An object s is referred to as random, and thus non-causal, if the algorithmic complexity $C(s)$ of s is about the length of s itself (in bits), i.e. it has no generating mechanism other than a print(s) function. Algorithmic complexity C is the accepted mathematical measure of intrinsic randomness of an object (independent of probability distributions), which is a generalization of statistical randomness and a refinement over the concept of Shannon entropy, as it does not depend on choice of probability distribution. Moreover, it has been proven to be mathematically robust (by virtue of the fact that independent definitions converge,[4,6,7] unlike the case of Shannon Entropy,[1] and because no computable measure can fully characterize (non-statistical) randomness (and therefore causality versus non-causality) due to the lack of universal computational power to test for every possible non-random feature[8]. C can also be seen as a measure of compressibility, but compression algorithms (e.g. LZ, LZW) are in fact entropy rate estimators and thus behave exactly like Shannon entropy (Fig. 1j), despite their generalized use as estimators of C.

The *Invariance theorem*[3,4,9] guarantees that complexity values will only diverge by a constant (e.g. the length of a computer compiler, i.e. a translating program between universal reference Turing machines $U_1$ and $U_2$) and will asymptotically converge. Formally, $|C(s)_{U1} - C(s)_{U2}| < c$.

$C(s)$ as a function that takes s to be the length in bits of the length of the shortest program p that generates s (and halts) is *lower semi-computable*, which means it can only be approximated from above. Proper introductions to the areas of finite algorithmic complexity and applications are provided in[2], and introductions to algorithmic (infinite sequence) randomness can be found in [5,10,11].

## 1.2 Algorithmic Probability

Algorithmic probability allows reaching a consensus of possible explanations of an underlying generating mechanism of a system (e.g. a network) at any time, thereby providing the most robust hypothesis for the available observable data. Algorithmic probability establishes and shows[15,16] that the consensus of several algorithmically likely solutions is the most likely one.

The chief advantage of algorithmic indices is that causal signals in a sequence may escape entropic measures if they do not contain statistical regularities, but they do not escape the metric of AP as there will be a Turing machine T capturing every statistical but also algorithmic aspect of s that compresses s but produces s in full with no more or less information than s itself (thus being lossless).

Let U denote a universal machine and let |p| denote the length of a program p. The Halting probability $\Omega$ [12] is the probability that U halts for random computer program p constructed bit by bit by random flips of a coin. That is,

$$\Omega_U = \Sigma_{p:\, T \text{ halts on } p}\ 2^{-|p|}$$

$\Omega$ is actually a family of probabilities as it depends on the enumeration of programs or reference universal Turing machine U, but optimal choices exist thanks to invariance-type theorems [13]. The Algorithmic Probability (AP) [9,13] (also known as Levin's semi-measure or Universal Distribution[14]) of a sequence s is the probability that s is produced by a computer program p, constructed bit by bit by flipping a coin, running on a reference universal Turing machine U divided by their Halting probability. Formally,

$$AP(s) = (1/\ \Omega_U)\ \Sigma_{p:T(p)=s}\ 2^{-|p|}$$

### 1.3 The Coding Theorem Method (CTM)

Lossless compression has traditionally been used to estimate the algorithmic content of an object s. The algorithmic complexity of a sequence s is then defined as the length of the shortest compressed file producing s when decompressing it (the file must contain the decompression instructions and thus always comes with a natural overhead). While lossless compression is an approximation of algorithmic complexity, actual implementations of lossless compression algorithms (e.g. Compress, Bzip2, gzip, PNG, etc) are based purely upon entropy rate,[1,17] and thus can only deal with statistical regularities of up to a window length size, hence being no more closely related to algorithmic complexity than entropy itself. Entropy and entropy rate, however, are not sufficiently sensitive and are not inherently invariant vis-a-vis object description[1,17]. AP, however, constitutes a true algorithmic approach to numerically estimating C(s) by way of the *algorithmic coding theorem* [Levin], formally relating these two seminal measures as follows:

$$C(s) = - \log AP(s) + O(1)$$

The *Coding Theorem Method* (or simply CTM)[15,16] is rooted in the relation between C(s) and AP(s), i.e. between the frequency of production of a sequence and its algorithmic probability. Unlike other computable measures, such as Shannon Entropy, CTM has the potential to identify regularities that are not merely statistical (e.g. a sequence such as 1234...) and that have as shortest program (and generating model) n := n + 1, that is, even sequences with high Entropy but no statistical regularities that are not random, have low algorithmic complexity and are thus causal as the result of an evolving computer program. As previously demonstrated[15,16], an exhaustive search can be carried out for a small-enough number of Turing machines for which the halting problem is known, thanks to the Busy Beaver game[18]. One strategy to minimize the impact of the choice of T is to average across a large set of different Turing machines, all of the same size[15,16]. Let (n, k) be the space of all n-state m-symbol Turing machines with n, k > 2. Then:

$$D(n, k)(s) = | \{T \text{ in } (n, k): T \text{ produces } s\} | / | \{T \text{ in } (n, k)\} |$$

is the function  assigned to every finite binary sequence s, where T is a standard Turing machine as defined in the Busy Beaver problem[18]. We remark that $0 < D(n, k)(s) < 1$, $D(n, k)(s)$, and is thus said to be a semi-measure, just as AP(s) is because the probability measure does not reach 1 due to non-halting machines. Then using the relation established by the Coding theorem [Eq 1], the measure of complexity which is heavily reliant upon AP used throughout this paper can therefore be defined[15,16] as follows:

$$CTM(s, n, k) = -\log_n(D(n, k)(s))$$

CTM is thus an upper bound estimation of algorithmic complexity[17]. For small values n and k, $D(n, k)$ is computable,[19,20] whereas for larger objects the estimation is based on an informed cutoff runtime based on both theoretical and numerical grounds, asymptotically capturing most of the halting Turing machines in polynomial time [5,21].

### 1.4 The Block Decomposition Method (BDM)

Because CTM is computationally very expensive (equivalent to the Busy Beaver problem), only the algorithmic complexity for short sequences (currently all sequences up to length k = 12) has thus far been estimated by the CTM method. To approximate the complexity of a longer sequence it is therefore necessary to aggregate the various computer programs that generate the string in a clever fashion by taking advantage of Shannon entropy. The new hybrid measure thus calculates local algorithmic complexity and global Shannon entropy at the same time. Formally, the BDM of a string or finite sequence s is as follows[22]:

$$BDM(s, l, n, k) = \sum_i^j CTM(x_i, n, k) \ + \ \log(s_i)$$

where $s_i$ is the multiplicity of $x_i$, and $x_i$  is the subsequence i after decomposition of s into subsequences $x_i$, of length l, with a possible sequence remainder y if $|y| < l$ and if its length is not a multiple of the decomposition length l. The parameter k runs from 1 to l that CTM can handle; m is an overlapping parameter to deal with the boundary conditions (the remainder sequence). The boundary conditions were studied in [22] where it is shown that BDM errors due to boundary conditions are convergent and vanish asymptotically, and that BDM is lower bounded by Shannon entropy and upper bounded by algorithmic complexity, thereby providing local estimations of algorithmic complexity and global estimations of entropy.

### 1.5 Graph Algorithmic Probability as Upper Bounds to Graph Randomness

We have shown that not all measures are robust vis-à-vis object description,[1] but that algorithmic probability and algorithmic complexity are, up to a constant term[23,26]. An adjacency matrix can thus be taken as the lossless description of a network as it is invariant (up to automorphisms). We look at the algorithmic probability of such a matrix being produced by chance by a computer program working on a grid (e.g. a so-called Turmite emulated by a Turing machine).

The algorithmic complexity K of a graph G is defined as follows[22,23,26]: Let A(G) be the adjacency matrix of G and Aut(G) its automorphism group, i.e. the set of  graphs of G isomorphic with itself. Then the algorithmic complexity of the graph is K(G) = min{K(A(G))|A(G) ∈ A(Aut(G))},

where A(Aut(G)) is the set of adjacency matrices for each G ∈ Aut(G). Since $K(D(G)) \sim K(A(G))$ for any computable lossless description D of $G^{26}$, we can safely write K(G), and it has been proven[22] that if G and G' are isomorphic graphs, then | K(G) − K(G') | < c, that is, G and G' have similar algorithmic information content. It has also been shown that numerical approximations that graphs in large automorphism groups have similar low algorithmic complexity and graphs in small automorphism groups can have both low and high complexity[23], thereby establishing a numerical relationship between algebraic complexity by group symmetry and algorithmic complexity approximated by BDM, with results conforming with the theoretical expectations. Let us call such an approximation of K(G) following BDM, C(G). The algorithmic complexity approximation C(G) of graph G is then defined by

$$C(G, x_i) = \sum_{(r_i, n_i)} x_i \log (s_i) + CTM(r_i)$$

where $x_i$ is composed of the pairs (r, n) with r an element of the decomposition of G in square sub-arrays of equal dimension and $s_i$ the multiplicity of each submatrix $x_i$ obtained by using 2-dimensional Turing machines in the calculation of CTM as introduced in [22,26]. In other words, we ask how often a random Turing machine can produce the adjacency matrix of G. We can see now how algorithmic complexity introduces a new dimension capturing the causal content of a graph by separating random-looking graphs from algorithmic random graphs:

There are clearly ER graphs that are not maximal algorithmic-random graphs. Consider p(t) to be a binary pseudo-random number generator with seed t. Let G be an ER graph of size n with edge density r. ER has rn(n − 1)/2 edges. Let every edge $e_i \in \{e_1,...,e_{rn(n-1)/2}\} \in G$ be connected to node vi ∈ $\{v_1,...,v_n\}$ if p(i) = 1 and disconnected if p(i') = 0. G is clearly ER but not maximally algorithmic-random because G is recursively generated by p with some seed t.

# 2. Methods

## 2.1 A Causal Perturbation Calculus as the Study of Algorithmic Change

At the core of the causal calculus is the estimation of the specific sequence of events—in the form of perturbations—that can change the fate and function of a system thereby ranking these perturbations and system's elements by the effects they can exert on the whole system. We will demonstrate that manipulating and reprogramming systems on the algorithmic-information space runs parallel to the dynamic state/attractor space through which a network or a system can be moved along optimized paths in different directions. Based on universal principles of the most powerful theory of induction and inference (namely algorithmic probability) in the sense that it is optimal (any other method is a special case) and drawing on recent numerical advances to produce estimations, we introduce a suite of parameter-free algorithms with the inherited property of great power to tackle the challenge of causal discovery, that is, to find and reveal generating mechanisms behind observations that may effectively control the dynamics of general non-linear systems removed from the limitations of classical statistical tools, without knowing the kinetic equations that require expensive numerical simulations, arbitrarily assuming non-linearity or requiring access to probability distributions.

The algorithmic causal calculus is based chiefly upon evaluating the algorithmic information of dynamical objects, such as strings or networks changing over time. Let S be a non-random binary file containing a string in binary:

**101010101010101010101010101010101010101010101010101010101010101010**

Clearly, S is algorithmically compressible, with $p_S$ = "Print(01) 35 times" a short computer program generating S. Let S' be equal to S except for a bitwise operation (bitwise NOT, or complement) in, say, position 24:

**10101010101010101010100010101010101010101010101010101010101010101010**

A short computer program that generates S' can be written in terms of $p_S$ that can already account for most of the string except for the perturbed bit, which $p_{S'}$ has thus to account for. A candidate program can be $p_{S'}$ ="Print(10) 11 times; Print(00); Print(10) 23 times". Clearly, the length in binary of the computer program $p_S$ generating S is upper bounded by $|p_{S'}|$, the length, in bits, of the computer program $p_{S'}$ generating S'. In other words, assuming shortest computer programs, we have C(S) < C(S') for any single-bit mutation of S, where C(S) = $|p_S|$ and C(S') = $|p_{S'}|$ are the lengths of the shortest computer programs generating S and S'.

Without loss of generalization, let S now be a binary file of the same size but consisting of, say, random data:

**0110110010001100110101000111011000110001101001110001000011001100011**

where S is now algorithmically incompressible. This means that S has no generating mechanism shorter than the string itself and can only be generated by a computer program of the form $p_S$ = Print(S), which is not much shorter than S itself. Let S' be the result of negating any bit in S just as we did for S and S'. Then $p_{S'}$ = Print(S'). There are 2 possible relations between $p_{S'}$ and $p_S$ after a single bit perturbation (bit negation), either $p_S < p_{S'}$ or $p_S > p_{S'}$ depending on S' moving towards or away with respect to C(S). However, perturbing only a single bit cannot result in a (much) less random S' because $p_S$ = Bitwise(Print(S'),n) can be used to reverse S' into S which is random, but both Bitwise and n (the bit index to be changed) are of fixed and small length and so do not contribute to the already high algorithmic complexity/randomness of the original random string (which can be of any length), contrary to the original assumption that S is algorithmically random (or not compressible).

As also illustrated in Fig. 1abc, this means that perturbations to an algorithmically random object have a lower impact on their generating mechanism (or lack thereof) than perturbations to non-random objects--- with respect to the generating mechanisms of their unperturbed states---and thus this effect can be exploited to estimate and infer the causal content of causal and non-causal objects. In an algorithmically random object, any change goes unnoticed because no perturbation can lead to a dramatic change of its (already high) algorithmic content. Real-world cases will move in an intermediate region between determinism and randomness (Fig. 7).

Let {S'} be the set of all possible mutations of S, and S'$_n$ an instance in {S'} with n from 1

to the power-set cardinality of $|S'|$ the length of S. Then each evaluation of $C(S'_n)$ represents the length of a possible generating mechanism accounting for each perturbation, intervention, past or future evolution of a string as a dynamic object changing from S to $S'_n$. These models and trajectories have the advantage of highlighting the principles by which a system or a network is organized, uncovering candidate mechanisms by which it may grow or develop. The causal calculus consists thus in studying the algorithmic-information dynamic properties of objects with a view to constructing an algorithmic-information landscape to identify and rank the elements by their algorithmic contribution and the changes they may exert on the original object, moving it towards or away from randomness. In what follows, we demonstrate that insights gleaned from the algorithmic information landscape can effectively be used to find and unveil the dynamics and reprogramming capabilities of the systems, starting from a reconstruction of space-time dynamics and their initial and boundary conditions, helping infer the (most algorithmically likely) generating mechanism of an evolving system from a set of (partial and even disordered) observations (see Fig. 3). We will explore the potential of this calculus to characterize genes in regulatory networks and to reprogram systems in general, including specific examples, theoretical and experimental, focusing on synthetic and biological data.

To date, there are no alternatives to applying non-linear interventions to complex systems in the phase space other than to make strong assumptions (e.g. linearity or mass probability distributions) to perform simulations of possible dynamical trajectories often requiring unavailable computing resources. This new calculus, however, requires much less information and makes significantly less assumptions to produce a collection of guiding causal interventions through desired even if rough dynamical trajectories that promises a wide range of applications.

## 2.2 Manipulating and Steering Systems

In practice, estimating the algorithmic complexity of a deterministic system $C(S_t)$ is characterized as *lower semi-computable,* meaning that it can only be approximated from above. Recent numerical advances—alternatives to lossless compression algorithms—have, however, led to estimations[4,5] based on the seminal concept of *algorithmic probability*[2,3] that goes beyond other methods, including traditional statistics, including Shannon Entropy and lossless compression that is better equipped to tackle causation,[6] This is because the estimation of $C(S_t)$ entails finding algorithmic models reproducing $S_t$ (or versions close to $S_t$) that can be aggregated to produce a set of candidate generative models of $S_t$. Such novel methods have found a wide range of applications in areas such as cognition[7,8], protein folding[24] and logic circuit design[25] to mention a few among several others. Here we take advantage of these fundamental and numerical advances to tackle a problem related to finding the mechanisms underlying the design and control of systems, in particular biological but first we explore the power of this calculus on simpler objects and discrete dynamical systems such as cellular automata.

**Table 1** The algorithmic information dynamics of a string (step 0) pushed towards and away from randomness by digit removal estimated by Algorithmic Probability (BDM, see Sup. Inf.). The initial string consists of a simple segment of ten 1s followed by a random-looking segment of 10 digits. The resulting strings mostly extract each of the respective segments.

| step | Towards randomness | Away from randomness |
|------|--------------------|----------------------|
| 0 | 11111111110101100011 | 11111111110101100011 |
| 1 | 1111111110101100011 | 1111111111101100011 |
| 2 | 111111111001100011 | 111111111111100011 |
| 3 | 1111111001100011 | 1111111111110011 |
| 4 | 1111111001100011 | 1111111111111011 |
| 5 | 111111001100011 | 111111111111111 |
| 6 | 11111100100011 | 1111111111111 |
| 7 | 1111100100011 | 111111111111 |
| 8 | 111100100011 | 11111111111 |
| 9 | 11100100011 | 11111111111 |
| 10 | 1100100011 | 1111111111 |

**Table 2** Neutral digit deletion maximizes the preservation of the elements contributing to the algorithmic information content of the original string thus the most important (computable) features (of which statistical regularities are a subset). Here applied to a string (step 0) after removal of 10 digits.

| step | string |
|------|--------|
| 0 | 1110101100011101111011111010 |
| 1 | 1110101100011101111011111010 |
| 2 | 1110101100011101111101111010 |
| 3 | 1110101100011011110110111010 |
| 4 | 1110101100011101111011011100 |
| 5 | 1110101100011101111101100 |
| 6 | 1110101100011101111101100 |
| 7 | 1110101100011101111011 |
| 8 | 111010110001110111101 |
| 9 | 11101011000111011110 |
| 10 | 1101011000111011110 |

     Tables 1 and 2 illustrate three kinds of algorithmic shifts that a calculus identifying the elements that can move an object towards and away from randomness can deliver. The concept is at the core of the causal algorithmic calculus as applied to, for example, strings before moving to more sophisticated and objects such as networks. Fig. 4abc demonstrates how the same calculus can be applied to networks to shift them towards or away from randomness just as we did for strings, and we show how these shifts impacts the possible dynamics of and on a network as demonstrated in Fig. 1abc illustrating connection. As illustrated, for a more random or simpler object such as a network the set of possible interventions that can act upon it have little to no effect. However, objects removed from simplicity and randomness have richer candidate generative models that are able to perform greater shifts under the effects of guided perturbations. Fig 1(e-j) demonstrates the advantages of moving from previous purely statistical approaches to algorithmic approaches more deeply related to causation and how an object such as a sequence of the positive integer numbers can be explained and assigned high causal

content while for statistical approaches such as Shannon Entropy it would appear random (if, as it is most of the time, there is little to no access to probability distributions and knowledge about the deterministic nature of the source).

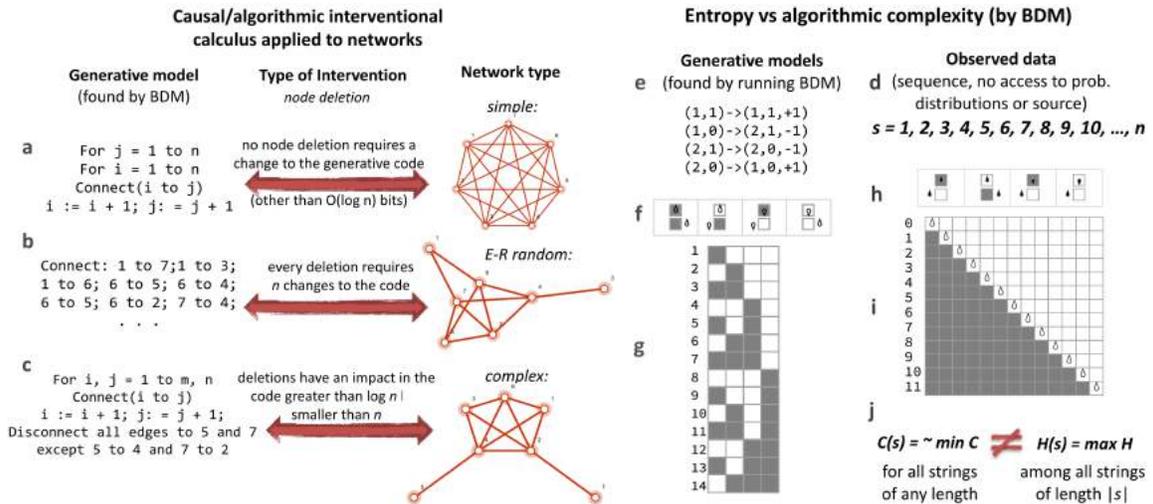

**Figure 1 From Statistical Correlation to Algorithmic Causation. (a,b,c)** Graphs of different origin require different encodings capable of recursively generate each graph, interventions to the graph may or may not have effect on the candidate algorithmic models depending on their algorithmic causal content. **(d)** A sequence such as *s* cannot be characterized by measures based on Entropy or classical statistics, but it can be characterized as of low algorithmic complexity because more than 1/3 of all possible Turing machines with 2 states encode a decimal counter and are thus many small computer programs that encode a highly algorithmic sequence that may not have any statistical regularity (called the Champernowne constant, *s* has been proven to be *Borel normal* and thus of maximal Shannon Entropy). **(f)** A computer program whose halting criterion is the leftmost head position of the Turing machine and **(g)** respective space-time evolution whose output in binary reproduce the sequence *s* in decimal **(h)** A unary non-halting computer program that computes *s* directly followed by **(i)** its space-time diagram effectively encoding the sequence generating function f(x) = x + 1. **(j)** It is thus clear that while Entropy H(s) can diverge from algorithmic (Kolmogorov-Chaitin) complexity C(s), it is C(s) encoding the simplicity of s and therefore characterizing its causal mechanistic nature by way of having been found by a procedure based on what will be described and exploited in this paper at the heart of the causal interventional calculus.

## 2.3 Networks as Computer Programs

Algorithmic Probability (AP)[2,3,9] deals with the challenge of inductive inference[10], and it is the obverse of the rigorous mathematical formalization of (algorithmic) randomness[1,3,9,11] by way of the so-called (algorithmic) *Coding theorem*[3] formally relating complexity C and probability AP. For example, the AP(G) of a causal graph or network G can be defined[12] as the probability that a random computer program constructed bit by bit–by, e.g., random flips of a coin–outputs a (lossless) description of G, d(G), e.g. its adjacency matrix. Because AP is *upper semi-computable*, it allows algorithmic complexity C to be approximated. Among the most remarkable properties of AP and C is that they cannot be refuted at an arbitrary significance level by any computable measure[3], and estimations of C(d(G)) asymptotically converge to C(G), independently of d, up to a relatively small constant[2,9,11] (Supplement Section 1, *invariance theorem*). The idea behind the numerical methods for estimating AP[4,5] is to enumerate the set of computer programs that explain and generate, in full or part, the generating system representing the predictive computational model of the observables (Supplement Section 1, *CTM* and *BDM* methods).

The most interesting application of this algorithmic calculus is to evolving systems. In a deterministic dynamical system S, the length of the shortest generating mechanism f describing a system's state (in binary) at time t, denoted by $C(S_t)$, can only grow by a function of t, more specifically $\log_2(t)$. This is because in a deterministic dynamic system, every state $s_{t+1}$ (encoding its own initial condition) can be calculated from $S_t$, i.e. $S_{t+1} = f(S,t)$. This trivial but fundamental property of deterministic dynamic systems can be exploited to find the set of perturbations of a system's state $S_t$ related to a set of perturbations $S_{t'}$ such that deviations from $\log_2(t)$ indicate non-causal trajectories and disconnected patches unrelated to the originally observed dynamic system. When a system is not completely isolated and some of its parts seem not to be explained by any other state of the system, thereby appearing non-deterministic, those patches can thus be exposed and identified as foreign to the system's normal cause in the algorithmic perturbation analysis.

Formally, the algorithmic calculus consists in the estimation of the change of algorithmic information content in a network G after an intervention. We define an element e in G to be *negative* if $C(G) - C(G\backslash e) < 0$, where $G\backslash e$ is a mutated network G without element e, moving G towards algorithmic randomness, *positive* if $C(G) - C(G\backslash e) > 0$, and *neutral* otherwise (Fig. 2a-e).

For example, a *maximally random network* (Fig. 2c) has only *positive* (blue) elements (Fig. 5c) because there exists no perturbation that can increase the randomness of the network either by removing a node or an edge, as it is already random (and thus non-causal). We denote by *information spectra*(G) (see Fig. 2f) the list of non-integer values quantifying the information-content contribution of every element (or subset of elements) of G and σ(G) the *signature* of G (Fig. 2g-i) which is the sorted version from largest to smallest value of the *information spectra*(G). For example, in a cycle graph, all nodes and edges have the same information content because any removal leads to a path graph, therefore the information *spectra(G)* is the set of unsorted values $\{C(G) - C(G\backslash e_i)\}$ for all *i* elements of G, while the σ*(G)* is the same set but as a sorted list $\{C(G) - C(G\backslash e_i)\}$ from greatest to lowest values. *spectra(G)* is informative to perform an ab-initio identification of, for example the vulnerable breaking points in regular S-W networks (Fig. 2k), whereas the removal of neutral elements (Fig. 2l) minimizes the loss of information relevant to the description of a network, if important, such as graph-theoretic properties (Fig. 2l) and its *information spectra* (by design), as it maximizes the preservation of the original algorithmic information content and thus represents an optimal parameter-free method for dimensionality reduction in the sense that it minimizes the loss of information and thus preserving the most important features of a system disregarding its nature (Fig. 2l-p, Sup. Inf. Section 2 & Methods). To save computational resources, in numerical experiments we will for the most part consider single perturbations only, rather than the full power-set of all of them.

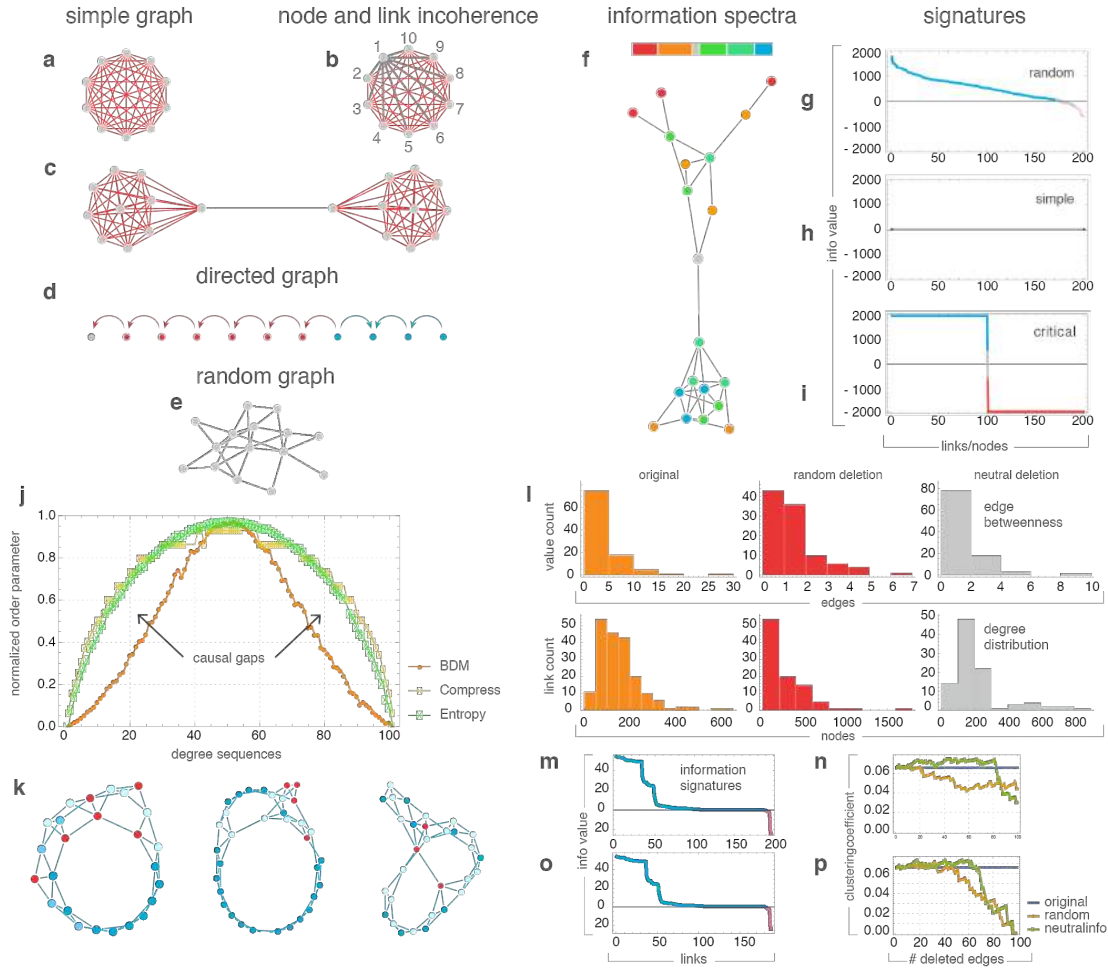

**Figure 2. Basic Concepts and Features (a)** Nodes and edges identified according to their information contribution to each network by evaluating the effect they have upon removal: red the element moves the network towards randomness, gray moves the network away from randomness by a logarithmic factor only, hence neutral. **(b)** *Neutral information set*: removal of node 1 is equivalent to the removal of all the edges connecting to node 1, even though all the individual edges are negative in themselves (we call this effect `incoherence'). **(c)** A random connection between two complete graphs is positive because its removal makes the generating mechanism of 2 complete graphs shorter than the 2 complete graphs randomly connected. **(d)** Information analysis on a directed graph identifies changes of direction. **(e)** All nodes and edges in a non-recursive random graph are neutral because no element can (significantly) move the network towards randomness (or simplicity). **(f)** Each element is ranked according to their algorithmic/causal contribution in an *information spectrum* identifying their contributions to the algorithmic model of the original network, elements coloured in red move the network towards randomness, elements towards blue move the network away from randomness, a network can be red or blue shifted depending on its causal content and the interconnection/dependence among all its elements. **(g,h,i)** The *signature* of a graph is the same *info spectrum* sorted from highest to lowest rank and is used to profile classes of networks. **(j)** Distribution of values of Entropy versus algorithmic complexity (BDM) of all strings of length 12 normalized by maximum Entropy. Some strings are less random than Entropy and lossless compression suggest. The gaps are the causality discovery gain by using algorithmic complexity

pinpointing cases in which strings may look statistically random but are causally not random. **(k)** The same techniques can pinpoint elements (inverse colouring than the *information spectrum*) at breaking points in evolving random graphs from regular graphs according to the Watts–Strogatz model. **(l)** Neutral node and neutral edge removal (i.e. those that do not move the network towards or away from randomness more than log(n) with n the size of the network, otherwise said those elements that preserve the information *signatures*) is able to preserve key graph-theoretic properties such as edge betweenness **(i top),** degree distribution **(I bottom)** and clustering coefficient **(n,p)**, and are thus optimal for dimensionality reduction (see MILS in the Sup. Inf.).

Since neutral elements do not contribute to the algorithmic content of a system they do not affect the length of the underlying generating program (see Fig.1abc and Fig.2l-p), which means that the network can recover those neutral elements at any moment by simply running the system back to the point when the elements were removed (Fig. 2a). This process of identification of algorithmic contributing elements allows systems to be 'peeled back' to their most likely causal origin, unveiling their generating principles (Supplement Section 2), which can then be used as a handle to causally steer a system (Fig. 2b,c) where other measures fail (Fig. 1e-j and Fig. 2d and Extended Data Fig. 2)[6]. Analogously, elements can be added to a network to increase or maximize its algorithmic information content, thus approximating a *Maximally Algorithmic-Random (MAR) graph* that can be used for maximum-entropy modelling purposes, with the advantage of discarding false maximal entropy instances (Supplement Section 2). In contrast to neutral elements, extreme (negatively or positively) valued network elements hold and drive the network towards or away from algorithmic randomness extending the current study of networks that has been constrained to mostly graph-theoretic, statistical or entropic approaches thereby adding another dimension of research (see Extended Data Fig. 1).

Observers (Fig. 3a) have only limited access to any system's generating mechanism denoted by P, or to the precise dynamics D of the same system. Each system's perturbation, or observation O(n), in time 0 to n, can correspond to an estimation of the complexity C(n) of O(n) based upon the likelihood of P explaining O(n). The objective when attempting to identify a system (or the consequences of system perturbations) is to access P by inspecting the system at observation intervals, capturing possible features to associate with P. Because knowledge about D, the dynamics governed by a system, e.g., ODEs or a discrete mapping such as a cellular automaton, is as a rule beyond reach, network-based approaches can conveniently focus on the relationships among a system's elements represented by a timeline T, thus serving as a topological projection of the system's dynamics based on, e.g., correlation (apparent row-column correlation from an observation in time).

Computer programs with empty inputs can encode both the dynamics and changing initial conditions of a system over time, constituting a true causal generating mechanism for all of a system's timelines (see Fig. 3a) where dynamics and topology are included. Not all systems are equally dependent on their internal kinetic dynamics. For example, network-rewriting systems updated according to replacement rules have no dynamics (Fig. 3d and Sup. Inf. Section 1 on Algorithmic Causal Reconstruction of Dynamic Systems).

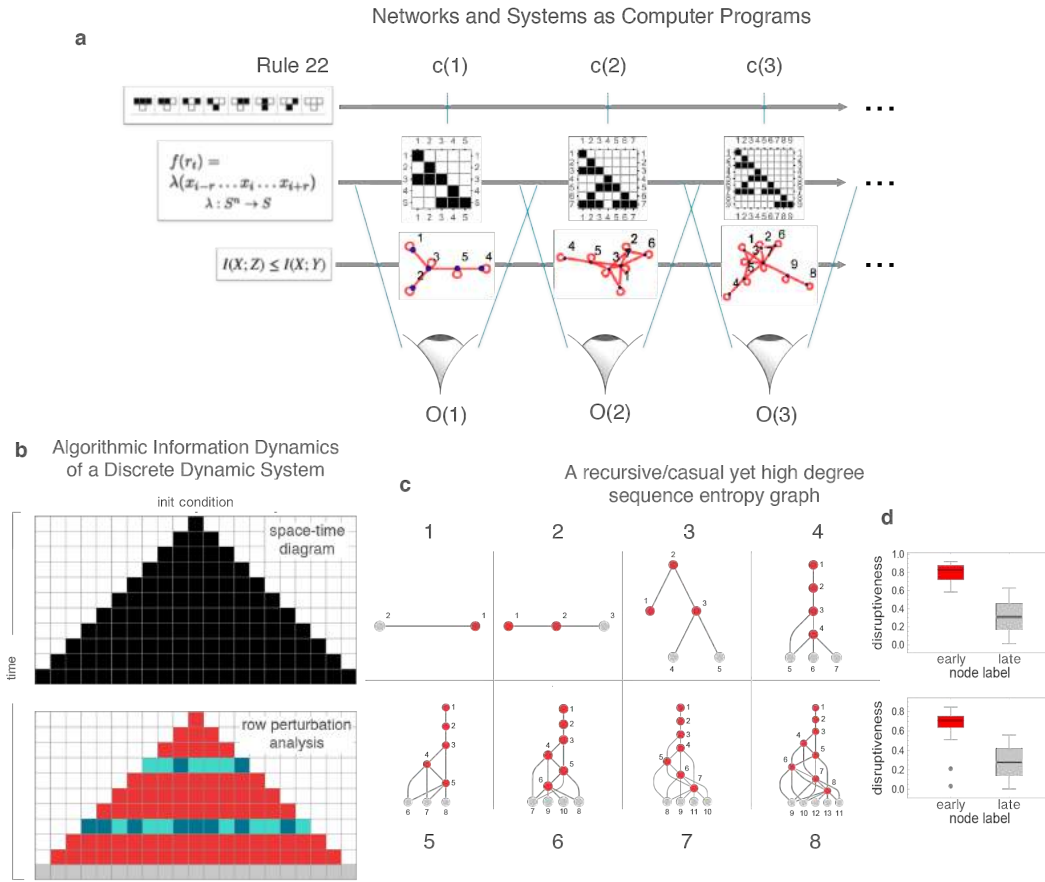

**Figure 3. Representations and Dynamical Systems (a)** The causal calculus being introduced can help reveal the generating mechanism of a discrete dynamical system, regardless of the different lossless representations it may have, that is, representations that preserve (most of) the information of the system from which it can be reconstructed in full. This ability comes from the property of closed deterministic systems that must preserve their algorithmic complexity along its time evolution given that its generating mechanism is always the same at every time step except for the time index that can be encoded by only log(n) bits. This means that any deviation from log(n) indicates that the system is not closed or not deterministic and is thus possibly interacting with some other system for which we can identify its interacting elements by perturbing them and measuring their deviation from log(n). **(b)** A one-dimensional evolving system displays the same information elements determining the different causal regions after an instantaneous observation following a perturbation analysis. In a Cellular Automaton, after 2 random row perturbations, the algorithmic calculus reveals which rows have been artificially perturbed, with grey cells showing the identified neutral row, the last (top-down) in the dynamic evolution, indicating the time direction of the system. See Fig. 4. **(c)** Unlike (a), Entropy is not invariant to different object descriptions. Shown here is a tree-like representation of a constructed causal network with low algorithmic randomness but near maximum Entropy degree sequence (SI 8), a contradiction, given the recursive nature of the graph and the zero Shannon entropy rate of its adjacency matrix, diverging from its expected Shannon entropy. **(d)** Latest nodes in the same graph depicted in (c) are identified by their neutral nature, revealing the time order and thereby exposing the generating mechanism of the recursive network (for details see Sup Inf 8 and 4).

### 2.4 Reprogrammability, a Measure of Algorithmic Change

From these first principles, where systems which are far from random, displaying an inherent regular structure, have relatively deeper attractors and are thus more robust in the face of stochastic perturbations, we derived a *(re)programmability index* according to which algorithmic causal perturbations of network elements pushing the system towards or away from algorithmic randomness reveal qualitative changes in the attractor landscape in the absence of a dynamical model of the system. A network is thus more *(re)programmable* if its elements can freely move the network towards and away from randomness. Formally, the *relative programmability* of a system G, $P_r(G)$ is defined by $P_r(G) := MAD(\sigma) / n$ or 0 if n = 0, where n := $max(|\sigma|)$ and MAD is the *median absolute deviation* (Supplement Section 1).

If $\sigma_N(G)$ are the elements that move G towards randomness, and $\sigma_P(G)$ the elements that move G away from randomness, then the *absolute programmability* $P_A(G)$ of G is defined as $P_A(G) := |S(\sigma_P(G)) - S(\sigma_N(G))| / m$, where m := $max(S(\sigma_P(G)), S(\sigma_N(G)))$ and S is an interpolation function. In both cases, the more removed from 0 the more reprogrammable, and the closer to 1 the less reprogrammable (Fig. 6). We then take as the *combined reprogrammability* of G the norm of the vector $||V_R(G)||$ on a *programmability space* given by the Cartesian product $P_r(G)$ x $P_A(G)$ (Supplement Section 1). These indices assign low values to simple and random systems and high values only to systems with non-trivial structures, and thus constitute what are known as a measure of sophistication that tells apart random and simple cases from highly structured, in this case quantifying the algorithmic plasticity and resilience of a system in the face of causal perturbations (Fig. 6 and Extended Data Fig. 3).

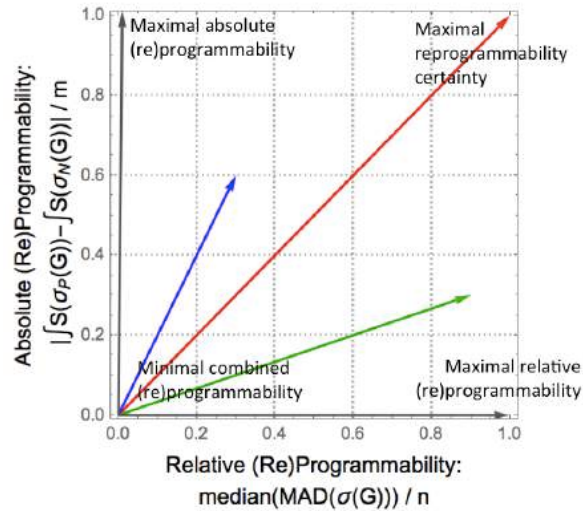

**Figure 6. Reprogrammability Space.** Illustration of the *(re)programmability space* defined by the Cartesian product $P_r(G)$ x $P_A(G)$, that is the (re)programmability indexes $P_r(G)$ and $P_A(G)$ in a 2-dimensional vector space. *Relative (re)programmability* $P_r(G)$ takes into account the sign of the different segments of $\sigma(G)$, i.e. how much of a signature segment below or above zero is not immune to small (convergent) numerical errors due to boundary conditions[1], whereas the *absolute (re)programmability* $P_A(G)$ accounts for the shape of the signature $\sigma(G)$, it measures the variability of the sample with robustness with regards to extreme values. (Supplement, section

1). Both of these indices contribute to the information about the (re)programmability capabilities of a system such as a network. The combined version is, effectively, a weighted index between the two (re)programmability measures that maximizes certainty measured by the magnitude of the vectors, where the closer to (1,1) or (re)programmability vector of magnitude $\sqrt{2}$, the greater the certainty of G being (re)programmed.

## 3. Results

### 3.1 Phase-space and mechanistic model reconstruction

In Fig. 4 it is demonstrated how the algorithmic calculus can help reconstruct discrete dynamical systems (illustrated using 1-dimensional cellular automata called Elementary Cellular Automata or ECA) with high accuracy from disordered states, and even index observations correctly, effectively providing a mechanistic generating model that can be run backwards and forwards in time. This is because late perturbations are more akin to a neutral information value (as established, in deterministic systems they should contribute at most log(n), with n representing the step index of the dynamical). The minor disagreements between the reconstructed order of observations shown in Fig. 4ab come from three sources: (1) it should be expected because more than one model may explain several arrangements of the same data, (2) we only apply single (all experiments reported in Fig. 4 except f and g) and double row perturbations (for Fig. 4f and 4g). Single perturbations have their limitations. For example, a greater number of simultaneous perturbations would be needed to correctly reconstruct deterministic order-R Markov systems for R > 1; and, finally, (3) algorithms to approximate the algorithmic information content are upper bounds and not exact values. The reconstruction of the same dynamical systems (cellular automata) taking into consideration an increasing number of observations illustrates (Fig. 4d-g), however, that the more data the more accurate the reconstruction thereby demonstrating that the numerical algorithm is not at fault and that reconstruction in practice is not only possible but computationally feasible and reliable. The method is scalable due to the clever shortcuts implemented in the BDM method (decomposition of causal patches that together can construct candidate models of much larger systems).

That we can reconstruct the space-time evolution of discrete dynamical systems with high accuracy from an instantaneous non-ordered set of observations (rows) demonstrates that we identify them as causal even among these random-looking systems such as ECA rules 73, 45 and 30, for which correlation values rho (Fig. 4) may be lower, though reconstructions are still qualitatively close. By exploiting the result that the later the step in time in a dynamical system such as an elementary cellular automaton the less disruptive the effect of the perturbation with respect to the algorithmic-information of the original system, we were able not only to reconstruct the cellular automata after row-scrambling but we gave each row a time index (Fig 4b). The automatic reconstruction of possible generating mechanisms by quantifying how disruptive a perturbation is to the algorithmic information content of the space-time evolution of a CA, allows us to extract the generating mechanism from the order in which perturbations are less to more disruptive in the hypothesized generating mechanism inferred from an instantaneous observation. Apparently simpler rules have simpler hypotheses, with an almost perfect correspondence in row order (Fig. 4a,b second columns from each pair). The ranking of the observations for some systems may look more random than others, but locally the relationship between single rows is mostly preserved, even among the more random-looking, either in the right or exact reverse order (indicating possible local reversibility).

The rule generating an observed deterministic dynamical system (such as an ECA) that maps $s_t$ to $s_{t+1}$ can then be derived from the causal deconvolution of blocks, as demonstrated in [21] and [22]. This is accomplished by looking at the smallest valid transformation among all consecutive observations in order to infer the influence of each event on the outcome where each local state $s_t$ leads to only a single future $s_{t+1}$ i.e. their `light cones'. This amounts to claiming that the regions are causally disconnected. The only assumption is that the system is deterministic, and that we can also infer the system's rulespace (e.g. the ECA), and thus the maximum length of the generating rule. But because we can always start by assuming the smallest possible rulespace size, the correct rule length will be the one that first causally disconnects all regions in a manner consistent with observations of the model 's rule. For example, to infer the rule behind ECA rule 254 (observations of the behaviour of which we could trivially rank and index, as we have also done for non-trivial cases) we would start from the simplest rule hypothesis that assigns every black cell to either black or white and vice versa. So we have: 1->1 and 0->0, or 1->0 and 0->1. However, neither of these cases is consistent with the observations, so we move on to consider a mapping of two cells to one, e.g. 0,1->1 and 0,0->1 and so on. However, 0,0 maps onto both 0 and 1, thereby failing to causally separate future states from the same past (invalidating the assumption that the system is deterministic/causal in nature). The rule that will first separate regions in a manner that accords with the observations is thus precisely ECA rule 254, that is, the triplets that are all sent to 1 (black cell), except 0,0,0->0. To properly rank observations by correct time index is thus a sufficient condition to make an optimal prediction based on the most likely generating rule/mapping---by the principle of the unnecessary multiplication of assumptions. More non-trivial cases of reconstruction and formalism for causal composition and decomposition may be found in [21] and [22].

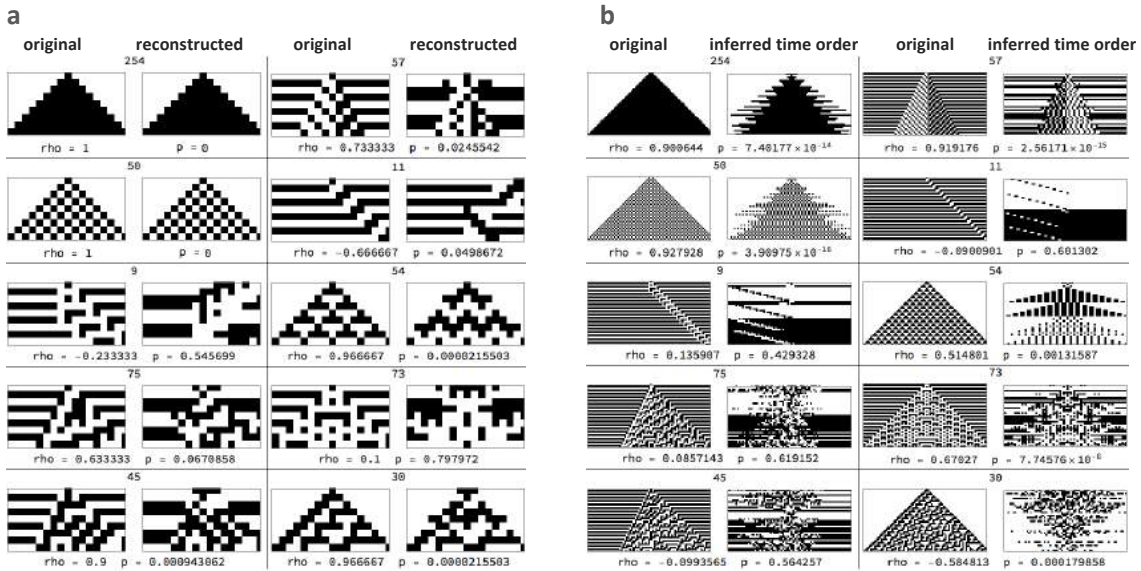

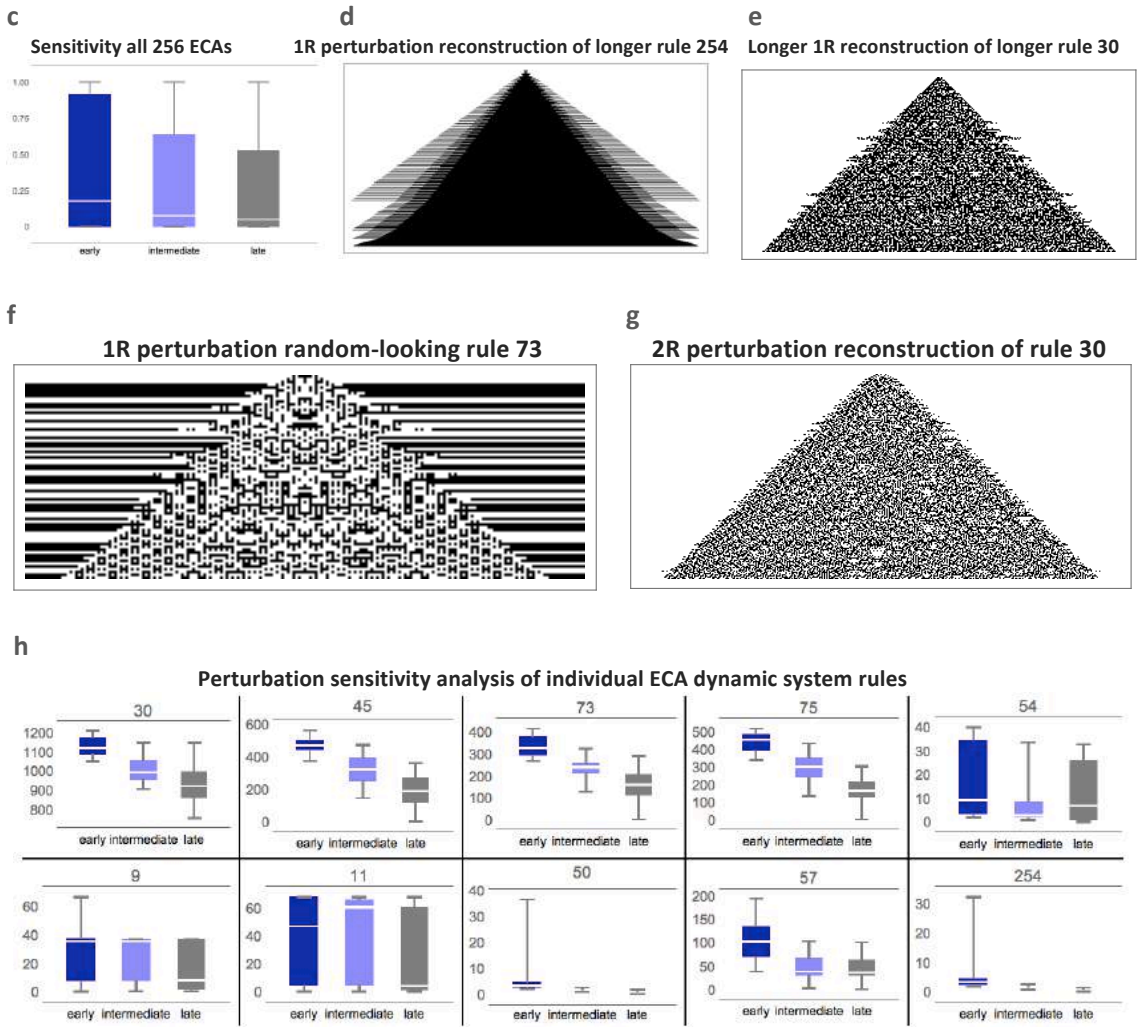

**Figure 4. Reconstructing Dynamical Systems. (a)** Reconstruction of the space-time evolution of dynamic systems (Elementary Cellular Automata or ECA[19]). Normal space-time evolution is displayed on the left-hand-side; on the right-hand-side are the reconstructed space-times after row scrambling by finding the lowest algorithmic complexity configuration among all possible 9! = 362880 row permutations (8 steps + initial configuration). All are followed by Spearman correlation values for row order. **(b)** Row time inference in linear time by generation of an algorithmic model that can run forwards and backwards, thus revealing the dynamics and first principles of the underlying dynamic systems without any brute force exploration or simulation. **(c)** As predicted, the later in time a perturbation is performed the less disruptive (change of hypothesized generating mechanism length after perturbation) compared to the length of the hypothesized generating mechanism of evolution of the original system. Each pair shows the statistical rho and p values between the reconstructed and original space-time evolutions, with some models separating the system into different apparent causal elements. **(d)** Depicted is the reconstruction of one of the simplest elementary cellular automata (rule 254) and **(e)** one of the most random-looking ECA, both after 280 steps, illustrating the perturbation-based algorithmic calculus for model generation in 2 opposite behavioural cases. **(f)** and **(g):** The accuracy of the reconstruction can be scaled and improved at the cost of greater computational resources by going beyond single row perturbation up to the power-set (all subsets). Depicted here are reconstructions of random-looking cellular automata (30 and 73 running for 200 steps) from single (1R) and double-row-knockout (2R) perturbation analysis. Errors inherited from the decomposition method (see Sup. Inf., BDM) look like 'shadows' and are explained (and can be counteracted) by numerical deviations from the boundary conditions in the estimation of BDM[20]. **(h)** Variations of the magnitude of the found effect are different in systems with different qualitative behaviour: the simpler, the less different the effects of deleterious perturbations at different times.

### 3.2 Algorithmic Information Dynamics of Networks

From a mathematical standpoint we have it that for every non-random network, there exists a generative (causal) program of a certain size (represented, e.g., by its degree sequence or any lossless matrix representation). In contrast, if a network has no shorter (lossless) description than itself, then it has no generative causal program and is defined as *algorithmically random*. This very generative program, and the number of possible halting states which it can be driven into, determine the number of attractors in a dynamical system. In a network with internal dynamics, both topological and kinetic details can be encoded in a full lossless description, and can thus be handled by the algorithmic causal calculus introduced here. Observations over time are the result of these two factors, but with no access to the generative program, deconvolution of all the measured elements contributing to the underlying system's dynamics is impossible, and we usually only keep a partial account of the system's dynamic output (see Fig. 3a and Supplement Section 2).

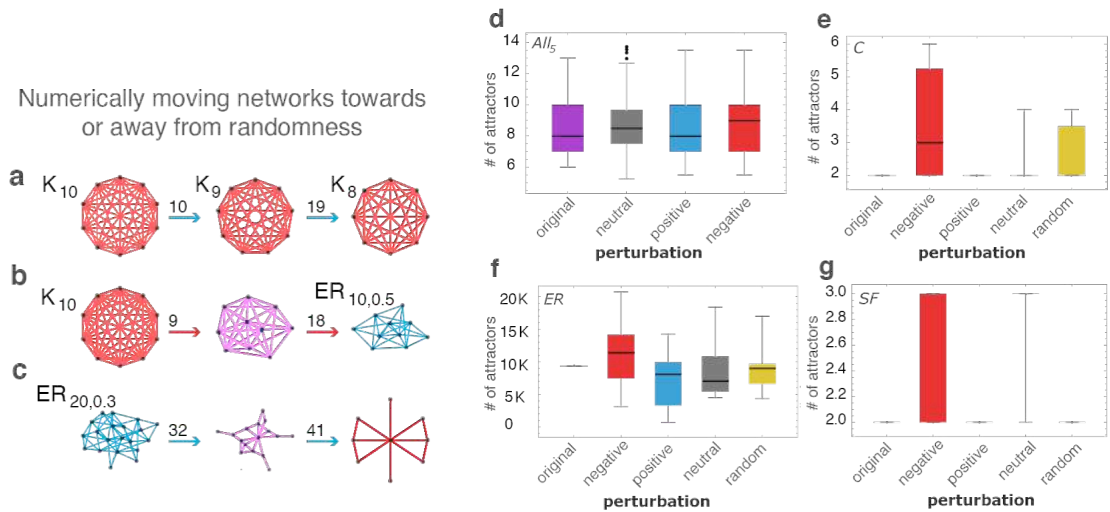

**Figure 5. Connecting Algorithmic Complexity and Dynamical Systems (a)** Numerically pushing a complete graph away from randomness by edge deletion produces complete graphs (after 10 and 19 steps respectively starting from the complete graph with 10 nodes $K_{10}$) as theoretically expected. **(b)** Pushing a network towards randomness, however, produces ER graphs approaching edge density ($d = 0.5$) also as theoretically expected. **(c)** Pushing a random graph towards simplicity reveals (after 32 and 41 steps) structured subgraphs contained in the original random one thereby revealing structure in randomness by single in silico perturbations. **(d)** Distributions (SI 6) of the number of attractors for all possible 5-node Boolean networks. The small difference is significant because the number of attractors in such small graphs is tightly

bounded. **(e)** Numerical calculation of the change in number of attractors in simple directed complete graphs, ER and scale-free networks converted into Boolean networks (SI 5 and 6). Scale-free networks, like regular networks, are more resilient in the face of perturbations.

This algorithmic calculus enables the identification of a system's causal core and facilitates the assessment of the causal contribution of a system's elements (detailed in Supplement Section 1 and 2). We evaluated whether this calculus could serve as a guide to reprogramming a system represented by a network corresponding to qualitative shifts in the *attractor landscape* associated with the system's behaviour, even in the absence of access to the dynamical system's equations. In low algorithmic content networks such as complete graphs, all nodes are immune to perturbations up to a logarithmic effect, leaving the basins of attraction and number of attractors the same (as a function of graph size only). *MAR graphs* (Supplement Section 2), however, have no (algorithmic or statistical) structure (by definition), and are thus predicted to have numerous shallow attractors. Moving an ER MAR network away from randomness will thus have an effect on the number and depth of its attractors, as it moves all the way away from randomness. Conversely, networks removed from randomness (e.g. a *simple directed regular* graph) have fewer but deeper attractors, but moving them towards randomness will eventually increase the number of attractors and decrease their average depth. These theoretical inferences are confirmed through simulation of Boolean networks (see Fig. 5d-g and Supplement Section 2). Based upon these principles, using, e.g., complete graphs as a model, we could predictively push networks towards and away from randomness (Fig. 5a-c). We also emulated Boolean dynamic networks with different topologies, predicting the nature and change in number of attractors (Extended Data Fig. 5 & Sup. Inf. Section 1) after pushing networks towards or away from algorithmic randomness. Here the number of attractors is based on the simulation of the networks equipped with random (or specific) Boolean functions (AND, OR, XOR, see Sup. Inf.) in their nodes and (randomly) directing the edges to be inputs and outputs, that is, a dynamics is mounted on each network and then the effect of the said perturbations studied with respect to both algorithmic randomness and the number of attractors of the network. The results show that not moving or moving towards and away from algorithmic randomness has a significant difference versus control experiments removing random nodes thereby establishing a connection between algorithmic complexity and dynamical systems. Consider that no alternative to perform educated perturbations existed to move a network along its attractor landscape to guide and steer its behavior other than to perform actual simulations or finding `drivers' using control theory making strong assumptions of linearity[23].

### 3.3 Application to Biological Networks and Reconstruction of Epigenetic Landscapes

We tested whether this algorithmic causal calculus can provide biological insight and has any explanatory power. First, we applied the calculus to an experimentally validated TF network of E-coli[13] (Fig. 6a). The negatively labelled genes (nodes) protect the network from becoming random and they were therefore found to be the genes that provide specialization to the cellular network, whereas positive nodes (genes) contribute to processes of homeostasis, pinpointing the elements of the network that make it prevail, since their removal would deprive the network of all its algorithmic content and thus of its most important properties. Then we analyzed a network controlling cell differentiation to assess the informative value of the qualitatively reconstructed attractor/differentiation landscape.

Proceeding from an undifferentiated cell state towards a more mature cell state, our calculus predicts fewer but deeper attractors in the differentiated state (Supplement Section 3). In Fig. 6b-d, we follow the process from a naive T cell differentiating into a Th17 cell signature[14]. This revealed an *information spectrum* with significantly different values over time and the *(re)programmability* (ratio of negative versus positive edges) was significantly higher in the first two time-points than in the final terminal time-point. Interestingly, the Th17 network signatures suggest information stability at the 48th point where only 3 nodes (STAT6, TCFEB and TRIM24) can further move the network towards greater randomness. After a gene enrichment analysis (Fig. 6b,c,d; Supplement Section 3, Extended Data Fig. 4), genes classified as having the most positive or negative information values comprised many genes known to be involved in T cell differentiation, such as transcription factors from the IRF and STAT families. Finally, retrieving network data from CellNet[15], we reconstructed heights in a corresponding epigenetic Waddington landscape for different cell types conforming to the biological developmental expectation (Fig. 6e; Supplement Section 3).

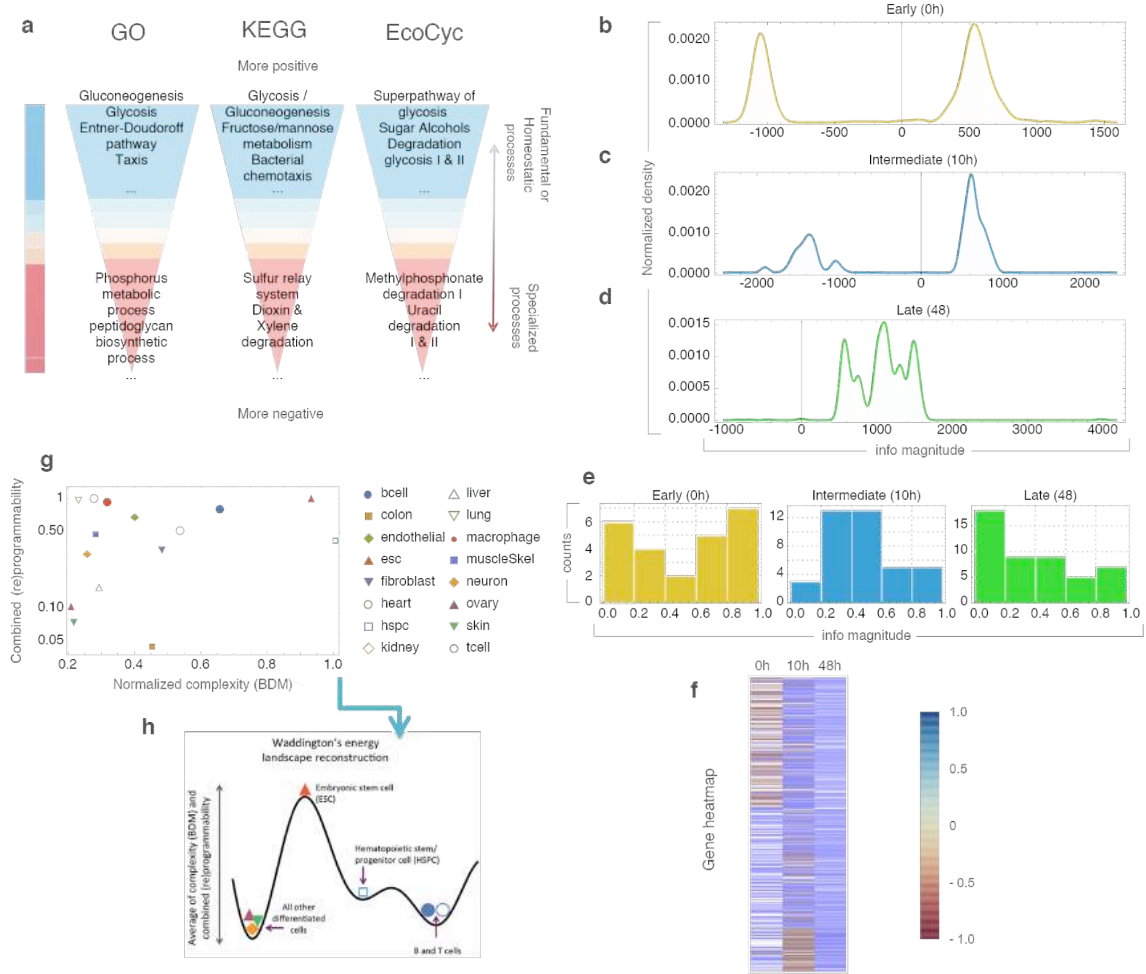

**Figure 6.** **(a)** K-medoid clustering of transcription factors by algorithmic-information node perturbation analysis on a validated e.coli network (extended figures 8-16) according to GO, KEGG and EcoCyc. Positive genes were found to be related to homeostasis while negative genes to processes of specialization. **(b-d)** Distribution of genes according to their causal information

value in the differentiation process from CD4+ to Th17 cells. **(e)** Uneven distribution of genes by information value strengthens the significance of the enrichment analysis (see Extended Figure 3). **(f)** Heatmap of normalized information values with approximately half the genes (Early) able to move the early network towards or away from randomness. Genes turn positive at the differentiated stage. **(g)** Charting the regulatory networks (for different cell-types from the Cellnet database using their complexity and *combined programmability*. (h) A sketch of the suggested epigenetic differentiation landscape reconstructed from the average of the reprogrammability and the algorithmic randomness (BDM) (see Sup. Inf. 1) for each cell network.

# 4. Conclusion

To summarize, the prevailing paradigm in system identification and control[16,17] can broadly be described as aiming to understand what the relevant features are in a system in order to formulate models to fit some properties of interest and then maximize the fitting of the model with respect to these properties. An unbiased identification of features is an NP complete problem, unless additional assumptions are made on the nature of the underlying data-distribution[18]. Thus despite advances in computational tools and fitting data—big or not—to a particular model, the issue of which the relevant properties are upon which to perform the model maximization or error minimization is unresolved. Since the causal calculus introduced here is based on fundamental mathematical results in algorithmic information theory, in combination with novel schemes for numerical evaluations, we have advanced a model-free proxy with which to estimate the qualitative shape of the dynamic possibilities of a system and thus make educated assumptions beyond current statistical approaches. Such an approach gives us a handle with which to intervene in and steer a system using these powerful parameter-free algorithms. Our results bridge concepts across disciplines and connect mature mathematical theories such as computability, algorithmic complexity and dynamic systems with the challenge of causality in science.

# Supplementary Information
## Section 1: Glossary of Terms, Concepts and Definitions

**Algorithmic causality:** The causal content c of a dynamical system $S_t$ running over t is given by the smallest c such that $C(S_t) \geq |S_t| - c$, where $| X |$ denotes the size of X. The difference $|S_t| - C(S_t)$ is an approximation of the causal content c of $S_t$. The causal content c of a non-causal system approximates log t, i.e. is very small, meaning that $C(S_t) \sim |S_t|$, and that the trajectory of $S_t$ is algorithmically random. For causal systems we have it that $C(S_t) - C(S_{t+1}) \sim \log_2 t$, i.e. the complexity of a causal system S is driven only by its evolution time t. All logarithms are in base 2 if not otherwise indicated.

**Algorithmic perturbation analysis**: Is the estimation of the effects of perturbations (e.g. by removal/knockout) of an element e (or set of elements) from S, denoted by S\e, on the original algorithmic information content C(S). Without loss of generalization, let's take as a system s, a network G = {V(G), E(G)}, a dynamic system, with V(G) a set of nodes and E(G) a set of links connecting nodes in V(G).

**Negative information element** (e.g. a node or edge): an element (or set) e in G such that: $C(G) - C(G\backslash e) < - \log_2 |V(G)|$ , i.e. the removal of e moves G towards randomness.

**Positive information element** (e.g. a node or edge): an element (or set) e in G such that: $C(G) - C(G\backslash e) > \log_2 |V(G)|$, i.e. the removal of e moves G away from randomness.

**Neutral information element** (e.g. a node or edge): an element (or set) e in G such that e is neither positive or negative:
$- \log_2 |V(G)| \leq C(G) - C(G\backslash e) \leq \log_2 |V(G)|$, where $|V(G)|$ is the size of the system, e.g. the vertex count of a network G.

**Algorithmic system inference of its generating mechanism:** Let s be a dynamical system, if $C(s_t) - C(s\backslash e) \sim \log_2 t$ we then call e a neutral perturbation. A perturbation e thus does not change the generating mechanism of s and $s_t$ can be recovered from $s_t\backslash e$ because $s_{t+1}\backslash e = s_t$. Otherwise e is disruptive (positive or negative), with degree of disruptiveness $C(s_t\backslash e) - C(s_t)$. In general, $C(s_t) - C(s_{t-n}) \sim n \log t$, providing the means to reverse a system in time and reveal its possible generating mechanism in the process. If the system is not reversible, a number of generating models may be formulated, thereby producing optimal hypotheses in the form of generative models.

***Spectra*(G)**: the list of all non-integer algorithmic-information contribution values of each element of G (e.g. edges or nodes, or both).

***Powerset spectra*(G)**: the list of all non-integer algorithmic-information values of each element in the powerset of elements of G (e.g. edges or nodes or both).

**Red shifted *spectra*(G)**: *spectra*(G) that contain more elements whose removal moves G more towards than away from randomness.

**Blue shifted *spectra*(G)**: *spectra*(G) that contain more elements whose removal moves G away from rather than towards randomness.

**σ(G)**: the information signature (or just signature) of G is *spectra*(G) list sorted from largest to smallest value. (see Extended Data Figures. 1-2).

**Δ(s):** the instantaneous programmability value of an element s in σ(G), indicating how fast or slowly s can move G towards or away from randomness. Formally,
$$\Delta(s) = |\ \sigma_i(G) - \sigma_{i-1}(G)\ /\ P(\sigma_i(G)) - P(\sigma_{i-1}(G))\ |.$$

**Incoherent information set:** a set whose individual elements or subsets have different information contribution values than the whole set.

**Coherent information set:** a set whose individual elements or subsets have the same information contribution value as the whole set.

***Information sensitivity***: the derivative of the absolute max value of the programmability of a graph in the (re)programmability curve (see Extended Data Figures 1-4), but numerically calculated by the rate of change of σ(G\e) versus σ(G) for all elements (or sets) e in G, i.e. the list of signatures for all e (or signature of signatures of G) capturing the non-linear effects of perturbations on G.

**MILS**: Minimal Information Loss Sparsification is a method to identify neutral elements that have zero or negligible algorithmic-information content value in a system or network, and can thus safely be removed, ensuring minimal information loss.

**MAR**: a Maximal Algorithmic Random graph (or system) G is an Erdős -Rényi (E-R) graph that is algorithmically random, i.e. whose shortest possible computer description is not (much) shorter than |E(G)|, where |E(G)| is the number of edges of G; or, |E(G)| − C(G) < c.

**1ˢᵗ Order randomness deficiency:** The algorithmic-information distance between a network/system and its algorithmically randomized version, e.g. a MAR graph for networks.

**2ⁿᵈ Order randomness deficiency:** The difference between information signatures by, e.g., Kolmogorov-Smirnoff distance, i.e. how removed a network is from its algorithmic (non-causal) randomization.

**Simply directed graph**: is the transformation of an undirected graph into a directed one such that the edge directions are chosen to minimize the number of independent paths and number of path collisions.

**MAD:** denotes the median absolute deviation, and is defined by:
$$\text{MAD} = \text{median}\ (|X_i - \text{median}(X)\ |).$$
MAD is a robust measure of the variability of a univariate sample.

**Relative (re)programmability**: $P_r(G) := \text{MAD}(\sigma(G)))\ /\ n$ or 0 if n = 0, where n = max($|\sigma(G)|$). This index measures the shape of $\sigma_P(G)$ and how it deviates from other distributions (e.g. uniform or normal).

**Absolute (re)programmability**: $P_A(G) := |S(\sigma P(G)) - S(\sigma N(G))| / m$, where $m := \max(S(\sigma P(G))$, $S(\sigma N(G)))$, where $m = \max(S(\sigma P(G))$, $S(\sigma N(G)))$ and $S$ is an interpolation function. This measure of reprogrammability captures not only the shape of $\sigma_P(G)$ but also the sign of $\sigma_P(G)$ above and below x = 0.

**Programmability landscape**: the Cartesian product $P_r(G) \times P_A(G)$.

**Combined (re)programmability**: $||V_R(G)|| = \sqrt{P_R^2(G) - P_A^2(G)} \leq \sqrt{2}$.
The combined reprogrammability is a metric induced by the norm $||V\_R(G)||$ defined by the Euclidean distance between two (re)programmability indices. This metric combines the relative and absolute (re)programmability indices, and takes into equal account both the sign of the signature $\sigma(G)$ of G and the shape of $\sigma(G)$, consequently minimizing the impact of uncertain sign estimations due to (convergent) errors in the calculation of algorithmic complexity[1] attributable to boundary conditions (see Graph Algorithmic Probability as Upper Bounds to Graph Randomness).

**Natural (re)programmability:** is the expected theoretical (re)programmability of a system or network, compared to its estimated (re)programmability, e.g. for a complete graph all nodes and all edges should have the same algorithmic-information contribution, and thus $\sigma(G)$ can be analytically derived (a flat uniform distribution x = log |V(G)| with |V(G)| the node count of G). Thus all the nodes of a complete graph are 'slightly' positive (or more precisely, neutral, if they are 'positive' by only log |V(G)|).

### Algorithmic-information Causal Interventional Calculus

The core of the causal calculus is based upon the change of complexity of a system subject to perturbations, particularly the direction (sign) and magnitude of the difference of algorithmic information content C between two graphs G and G', e.g. the removal of e from G (denoted by G\e). The difference | C(G) − C(G\e) | (see Supplement, Section 1) is an estimation of the shared algorithmic mutual information[1] of G and G\e. If e does not contribute to the description of G, then | C(G) − C(G\e) | ~ $\log_2$|V(G)|, where |V(G)| is the node count of G, i.e. the difference will be very small and at most a function of the graph size and thus C(G) and C(G\e) have almost the same complexity. If, however, | C(G) − C(G\e) | < $\log_2$|V(G)| bits, then G and G\e share at least n bits of algorithmic information in element e, and the removal of e results in a loss of information. In contrast, if C(G) − C(G\e) > n, then e cannot be explained by G alone nor is it algorithmically not contained/derived from G, and it is therefore a fundamental part of the description of G with e as a generative causal mechanism in G, or else it is not part of G but has to be explained independently, e.g. as noise. Whether it is noise or part of the generating mechanism of G depends on the relative magnitude of n with respect to C(G) and to the original causal content of G itself. If G is random then the effect of e will be small in either case, but if G is richly causal and has a very small generating program, then e as noise will have a greater impact on G than would removing e from the description of an already short description of G. However, if | C(G) − C(G\e) | ≤ $\log_2$ |V(G)|, where |V(G)| is the vertex count of G, then e is contained in the algorithmic description of G and can be recovered from G itself (e.g. by running the program from a previous step until it produces G with e from G\e).

For example, in a complete graph $K_{10}$ (Fig. 1a,b), the removal of any single node leads to a logarithmic reduction in its algorithmic complexity, but the removal of any single edge leads to an increase of randomness. The former because the result is simply another complete graph of a smaller size, and the latter because the deleted link would need to be described after the description of the complete graph itself. However, the removal of node 1 (Fig. 1 b) is equivalent to the removal of the set of all edges connecting to node 1, so the set of all these edges is a positive information set, even though all its individual edges are negative, a nonlinear phenomenon that we call *information incoherence*. Connecting two complete graphs at a random node (Figure 1c) designates the connecting link as positive because its removal pushes the network towards simplicity, the minimal description of 2 $K_{10}$ graphs being shorter than the minimal description of 2 $K_{10}$ graphs plus the description of the missing link at random points. Such a link can also be seen as an element disconnecting 2 networks, hence a network of networks. Its identification and removal would thus reveal the separation between two networks. In general, positive elements will identify the major structures generated by the most likely (and simplest) generating mechanism given the observation, and odd elements will stand out as negative, thereby identifying layers of networks that are independent of separable generating mechanisms, even removing apparent noise (external information) from the signal (the system's natural evolution) when such networks are richly causal. Random graphs are node- and edge-*blueshifted* (see Fig. 1g and Supplementary Information Glossary Section 1); simple graphs such as complete or wheel graphs are edge-*redshifted*. Perturbing (e.g. knocking-out) a node and recalculating the spectra changes the original spectrum in what is clearly a non-reductionistic approach to characterizing networks. All the methods introduced here also work on directed (e.g. Fig. 1d) and weighted graphs without any loss of generality.

Real-world networks as generated by physical laws are recursive according to classical mechanics (deterministic and reversible), and are thus on the left side in the schematic Extended Data Figure. 1, but they may also contain information about other interacting systems or be captured in a transient state that incorporates external signals pushing the networks towards randomness. We have quantified this concept by proposing different (Re) Programmability indices (see Supplement Section 1). Extended Data Figure. 1 summarizes some of the theoretical expectations and numerical results. There is a thermodynamic argument as to why the curve is negatively skewed: while it is easy and fast to move regular networks towards randomness as a function of the number of edges—there being about |E(G)| ways to move the network towards randomness such that the description of G moves to |G| + |e|, i.e. the description of, say, an edge removed, where |E(G)| is the edge count of G—there are far fewer ways to move a random network away from randomness. A MAR graph, for example, cannot be moved by edge or node deletion more than log |E(G)|. The result is compatible with the asymmetries in energy landscapes between moving systems towards fewer future attractors versus moving them back to states of a greater number of attractors, the latter requiring much more energy than the former.

**Minimal Information Loss Sparsification (MILS)**
Our causal algorithmic calculus defines an optimal parameter-free dimension reduction algorithm, which minimizes information loss while reducing the size of the original (network) object. The Minimal Information Loss Sparsification (or MILS) method is based on removing neutral elements while preserving the information content of a network, and therefore its properties, and it can be used for reduction by minimizing the loss of any informational feature of G that needs to be described and cannot be compressed into some shorter description of G (see Supplement Section 2 for the pseudocode and evaluation).

**Maximal Algorithmic Randomness Preferential Attachment (MARPA) algorithm**

The Maximal Algorithmic Randomness Preferential Attachment (MARPA) algorithm (MARP) (see Supplement Section 2 for the pseudocode and evaluation) can be viewed as a reverse algorithm in comparison to MILS. MARPA seeks to maximize the information content of a graph G by adding new edges (or nodes) at every step. The process approximates a network of a given size that has the largest possible algorithmic randomness and is also an Erdős-Rényi (ER) graph. An approximation of a 'Maximal' Algorithmic- Random (MAR) graph can be produced as a reference object whose generating program is not smaller than the network (data) itself and can better serve in maximum (algorithmic-) entropy modelling. See Supplement Section 1 for the proof of the existence of ER graphs that are not maximal algorithmic-random graphs

**Dynamical simulations using Boolean networks**

A Boolean network consists of a discrete set of Boolean variables each of which has a Boolean function (here, always the same for each node), which takes inputs from a subset of these variables. We conducted a first experiment on single-node and single-edge deletion effects on all possible Boolean networks with up to size n = 5 nodes, and with XOR, AND, and OR as Boolean functions. The output of a Boolean network is the state of the numbered sequence of states of its nodes. In a Boolean model in which a network is represented by a set of n Boolean variables, either Off (0) or On (1), the number of attractors cannot exceed $2^{n^{2^n}}$.

In general, in a connected network, each node is controlled by a subset of other nodes. The size of the controlling subset for each network depends on the connectivity pattern in the network [3, 4]. For example, in an E-R random graph with edges equally distributed with edge density p, if we change the state of any arbitrary node in the initial state, the effect on the dynamics of a network should be about the same on average, and this means the basin of attraction remains mostly unchanged. If the basin of attraction is of size M, the number of attractors is $(2^n)/M$. The size of M will depend on the network density p with $M \ll 2^n$. However, in a simply connected complete graph (minimizing edge collision c.f. Sup. Inf. Glossary), all nodes control all other nodes and there is only one attractor with basin of attraction size $2^n$. In modular scale-free networks, not all edges are statistically equally distributed and only a few nodes control many others, unlike an E-R random network, and they have significantly greater basin of attraction sizes and therefore a smaller number of attractors[2-4].

We estimated the algorithmic-information contribution of every node n (and every edge e) over all possible 33,554,432 5-node graphs. The estimation of the algorithmic-information contribution (see Supplement Section 1)) considered all vertices in the same orbit of the automorphism group of G, Aut(G), and the min of the information value C(G\n) with respect to the largest component of G according to the unlabelled definition of algorithmic complexity for unlabelled graphs[26 (main text)], thus correcting minor deviations of estimations of the complexity of C(G\n) by BDM due to boundary conditions[20 (main text)]. The calculation of C(G') for every G' in Aut(G) is, however, not feasible in general, as the production of Aut(G) and thus the calculation of C(G') for all G' in Aut(G) is believed to be in NP, thereby making the brute force exploration computationally intractable. However, it has also been shown that estimations of K(G) are similar to K(Aut(G))) up to a constant (the size of the graph generating program)[26 (main text)].

We performed the same edge perturbation experiments, removing all edges, one at a time, from larger graphs, [Fig3e] and comparing with state-of-the-art algorithms[5] the largest eigenvalue, number of different eigenvalues and number of attractors on the largest remaining connected component of the larger graphs. The experiment was repeated with Boolean functions AND, OR and XOR leading to the same results.

One can then apply uninformed perturbations to move networks towards statistical randomness based on this algorithmic-information calculus, and in a controlled fashion towards and away from algorithmic randomness, thus taking into account non-statistical and non-linear effects of the system as a generating mechanism, providing a sequence of causal interventions to move networks and systems at the level of the (hypothesized) generating model in order to reveal first principles and to control the side effects of such a system's manipulation at every step.

Random versus regular networks are sensitive in different ways. While an algorithmic- random network is hard to move fast along its algorithmic -random location (Extended Data Fig. 1-4), other changes in simple regular graphs have more dramatic effects (Fig1a v Fig1c), displaying different degrees of linear v. non-linear behaviour for different perturbations. In low algorithmic-content networks such as *simply directed complete graphs*, all nodes are immune to perturbations, leaving the basins of attraction and number of attractors the same (only proportional to their new size). From these principles, it is evident that systems that are far from random display inherent regular properties, and are thus more robust in the face of random perturbations because they have deeper attractors (See Supplement Section 2).

**Algorithmic Causal Reconstruction of Dynamic Systems**
The theory of algorithmic complexity provides means to find mechanistic causes through most likely (simplest) algorithmic models, helping to reverse engineer partial observations from dynamic systems and networks.
The causal reconstruction method of a system (e.g. a network or cellular automaton) M is as follows:

1) Estimate the information contribution of every element e in O(n), the sequence of instantaneous observations O from time 0 to n.
2) The set of neutral elements {e} is the set of those elements whose algorithmic-information content contribution to the complexity O(n) is of a logarithmic nature only with respect to C(n).
3) Remove neutral elements {e} from O(n) and repeat (1) with reassigned O(n) := O(n)\{e}.
4) After m iterations the reverse sequence of observations O(n)\{e} provides an indication of the evolution of the system in time, thereby yielding a hypothesis about the generating mechanism P producing O(n) for any n, and unveiling the initial condition in the last element of the above iteration, or the first after reversing it (see supplement, section 2 for more details and an example).

## SI SECTION 1 REFERENCES

# Section 2: Parameter-free Algorithms: pseudo-codes and evaluations

**Dynamical simulations by Boolean networks**

We explored whether the algorithmic content, or more precisely the information spectrum, of a system/network, influences transitions between different stable states, thereby effectively providing a tool with which to steer and reprogram networks. We observed an average decrease in the size of reachable states for all nodes (mean value), and the distribution of reachable states becomes more clustered (standard deviation), and more symmetrical (skewness) for all graphs with 5 nodes and single deletion. Positive info nodes had a similar effect as the deletion of a hub in the network. Absolute and relative negative nodes have a similar effect, whereas neutral (no information change) nodes preserve the distribution skewness closest to the original.

Histograms of perturbation effects on all graphs of size 5 nodes using functions XOR, AND, and OR produced similar results (see Fig3d & raw data infoedgesmotifs5.csv). Due to the small size of the graph we were able to control for graph automorphisms to correct minor BDM errors produced by boundary conditions[20 (main text)]. Two objects x, y are in the same orbit if there is an automorphism $\varphi$ in Aut(G) such that $\varphi(X) = Y$ (equivalently, $X = \varphi^{-1}(Y)$). In the algorithmic perturbation analysis, if elements $e_1$, …, $e_n$ in E are in the same orbit in Aut(G) we take the perturbation of every element in E to be equal to min{| $C(G\backslash e_1) - C(G)$ |, … , | $C(G\backslash e_n) - C(G)$ |, …}. In other words, the effect of every element $e_i$ in E on G is the same. The automorphism group Aut(G) was generated with the help of public software[1,2] for this experiment. For larger networks, however, this becomes computationally expensive, in the context of the perturbation analysis, and thus, because we have shown that K(G) ~ K(Aut(G))[26 (main text)], we continued calculating C(G) only.

In the exhaustive experiment over all connected graphs of node count 5, deleting the largest versus smallest node degree produced statistical differences as expected and previously suggested[4]. More relevant to our purposes, it was found that positive versus negative versus neutral information node/edge removal led to statistically different effects when executed in connected networks. Negative information node removal was interestingly not similar to lowest degree removal, yet significantly different statistically from control (random) node removal. Absolute and relative negative information removal had similar effects, and neutral (no information change) nodes/edges kept the distribution skewness closest to the original distribution, in accordance with the theory. For negative edges, the number of attractors was significantly increased (Fig5d), as the theory predicted.

**Minimal Information Loss Sparsification (MILS)**

Below, we provide the pseudo-code for the MILS algorithm. MILS allows dimensionality reduction of a graph (or any object) by deletion of neutral elements, thus maximizing preservation of the most important properties of an object as the algorithmic information content is invariant under neutral node perturbation. Let G be a graph. Then:

1. Calculate the *powerset spectra*(G) and let $E_j$ be the subset j in the set of all non-empty proper subsets of edges {$e_1$, . . . , $e_n$} in G.

2. Remove the subset $E_j$ such that $C(G\backslash E_j) < |C(G\backslash E_i)|$ for all $E_i$ in *powerset spectra*(G) (see

Glossary section 1), where $|C|$ is the absolute value of C.

3. Repeat 1 such that $G := G \backslash E_j$ until final target size is reached.

The algorithm time complexity class is in $O(2^{p(n)})$ (if there are no subsets with the same information value) because of the combinatorial explosion of the power set, but a more efficient suboptimal version of MILS iterates only over singletons:

1. Calculate $G \backslash e_i$ for all $i \in \{e_1,...,e_n\}$ i.e. *spectra*(G).
2. Remove edge $e_j$ in *spectra*(G) such that $C(G \backslash e_j) < |C(G \backslash e_i)|$.
3. Repeat 1 with $G := G \backslash e_j$ until final target size is reached.

We call $e_j$ a neutral information edge because it is the edge that contributes less information content (in particular, it minimizes information loss or introduces spurious information) to the network according to the information difference when removed from the original network.

Assuming that the estimations of C(G) and *spectra*(G) are definite and fixed (in reality one can always find tighter upper bounds, though, due to C's semi-computability), and MILS is a deterministic algorithm. Let G be a network and $i(e) = C(G) - C(G \backslash e)$ be the information value of element e in G with respect to G. If $i(e') > i(e)$ then MILS algorithm removes e first (by definition) because it minimizes the loss of information if the choice is to remove either e or e'. Thus we have it that $C(G \backslash e_1) = C(G \backslash e_2)$ if, and only if, $i(e_1) = i(e_2)$. However, it does not hold in general that $C(G \backslash e_1 \backslash e_2) = C(G \backslash e_2 \backslash e_1)$, that is, the removal of $e_1$ followed by the removal of $e_2$ from G, is not equal to the removal of $e_2$ followed by the removal of $e_1$ from G, even for $i(e_1) = i(e_2)$, because of non-linear effects (i.e. the removal of $e_i$ may modify the information contribution of all other $e_j$ in $G \backslash e_i$). This suggests that the only way to deal with these cases for MILS to be deterministic is the simultaneous removal of the set of elements $\{e_1, \dots e_n\}$ such that $i(e_1) = \dots = i(e_n)$. The time complexity of MILS thus ranges between the original $O(n^2)$ in the worst case to $O(1)$, when all nodes have the same information value/contribution to G and are thus removed in a single step. Therefore, set removal turns MILS into a proper deterministic algorithm that yields the same object for any run of MILS over an object G. Because any property of a network ultimately contributes to its information content (the amount of information to describe it), information minimization will preserve any potential measure of interest. We show in the following section that minimizing loss of information maximizes the preservation of graph theoretic properties of networks such as edge and node betweenness, clustering coefficient, graph distance, degree distribution and finally information content itself.

**Experimental evaluation of MILS using real-world networks**

Depicted in Fig2(i) and (j), is an example of a scale free network with 100 nodes and its original *information signature* (see Supplement section 1), which after neutral edge removal preserves the information signature (by design) after deleting 30 neutral edges, but also preserves graph theoretic properties such as edge betweenness, clustering coefficient and node degree distribution after deleting all graph edges, thus being superior to several other common sparsification methods (validated on 20 other gold-standard networks, [5] see also Supplement Section 1). This is because an element that is deleted will lead to a reduction of a network's algorithmic information content, so in the maximization attempt to preserve its algorithmic information content, only the less informative or most redundant properties of a network/system will be removed.

When MILS is applied to a set of well-known networks used before in pioneering studies, [5] we find that not only is the loss of information signatures and thus the non-linear algorithmic information content of the system minimized, but also that all the tested and most common graph-theoretic measures are maximally preserved. This contrasts with the application of random deletion and other common dimension reduction methods.

**Maximal Algorithmic Randomness Preferential Attachment (MARPA) algorithm**

MARPA allows constructions of a maximally random graph (or any object) by filling out the blanks, i.e. adding edges, for any given graph in such a manner that randomness increases. Let G be a network and C(e) the information value of e with respect to G such that $C(G) - C(G\backslash e) = n$. Let P = {$p_1, p_2, ..., p_n$} the set of all possible perturbations. P is finite and bounded by $P < 2^{|E(G)|}$ where E(G) is the set of all elements of G, e.g. all edges of a network G. We can find the set of perturbations e' in P such that $C(G) - C(G\backslash e') = n'$ with $n' < n$. As we iterate over all e in G and apply the perturbations that make $n' < n$, for all e, we go through all $2^{|E(G)|}$ possible perturbations (one can start with all |E(G)| single perturbations only) maximizing the complexity of G' = max{G | $C(G) - C(G\backslash e)$ = max among all p in P and e in G}. Alternatively, there is a configuration of all edges in G that maximizes the algorithmic randomness of G. Let such a maximal complexity be denoted by maxC(G). Then we find the sequential set of perturbations {P} such that maxC(G) - C(G) = 0, where C(G) - maxC(G) is a measure, related to randomness deficiency,[6,7] of how removed G is from its (algorithmic-) randomized version maxC(G) (notice that C(G) is upper bounded by maxC(G), and so the difference is always positive). Fig. 3a-c, shows how we numerically (single-element wise) moved a regular network towards randomness (in particular an E-R graph). Notice that while an ER network with edge density 0.5 is of maximal entropy, it can be of high or low algorithmic randomness, i.e. recursively generated or not,[1] but a high algorithmic-random graph is also ER because, if not, then by contradiction it would be statistically compressible and thus non-algorithmic random, because a graph with any statistical regularity cannot also be an algorithmic-random or an ER graph. One can also consider the absolute maximum algorithmic-random graph, denoted by amaxC(G) and disconnected from the number of elements of G (thus not a randomization of G), that is, the graph comprising the same number of nodes, but an edge arrangement such that $C(G) < C(amaxC(G)) \leq 2^k$ where k = (|E(G)|(|E(G)|-1))/4 is the maximum number of edges in G divided by 2 (at edge density 0.5 it reaches max algorithmic randomness. The process approximates a network of a size that has the greatest possible algorithmic randomness and is also an Erdős-Rényi (ER) graph. The pseudo-code is as follows:

1. Start with graph G (can be empty).
2. Attach edge $e_j$ to edge $e_j'$ in G such that $C(G \cup e_j') > C(G)$.
3. Repeat 1 with G := G $\cup$ $e_j'$ until final target size of graph is reached.

Generating a MAR graph is computationally very expensive with exponential time complexity in $O(2^{n^2})$ because at every step all possible attachments have to be tested and evaluated (i.e. all possible permutations of the adjacency matrix of size n×n), but small MAR graphs are computationally feasible, and they represent approximations of "perfect" ER random graphs, but unlike some ER graphs they cannot, in principle, be recursively generated with small computer programs. The intuition behind the construction of a MAR graph is that the shortest computer program (measured in bits) that can produce the adjacency matrix of the MAR graph, is of about the size of the adjacency matrix and not significantly shorter. Thus it can in some strict sense be considered the perfect ER graph. Every time that a larger graph, and therefore the addition of new edges, is needed, the computer program that generates it grows proportionally to the size of the adjacency matrix (See Supplement section 1 for algorithm and more details).

**Algorithmic Causal Reconstruction of Dynamic Systems**

There are systems whose internal kinetics fully determine the system's behaviour, i.e. attractor structure, such as Hopfield networks[8] and Boltzmann machines[9], which is independent of their fixed topology (complete graphs). Other networks are, however, more dependent on topology or geometry (e.g. disease networks or geographical communication networks). Boolean networks are governed both by their topological and internal kinetic properties as encoded by the connectivity of the node with the assigned Boolean function[10,11] to that very node. Each observation of a system is necessarily only a partial snapshot of the system's trajectory in phase space and it reveals only certain aspects of the generating cause, yet without any loss of generality one can use the causal calculus introduced here either on T, D or a combination of T and D in order to produce algorithmic models of causal generating mechanisms approaching P and producing T and D (see Fig.3 in main text). While the focus of the causal calculus introduced here is on T, it can readily incorporate D by moving to the phase space without any essential modification. We have included some examples using discrete dynamic systems such as cellular automata to show how the same calculus can be utilized. The same elements in D that move a system towards, or away, from randomness are, conversely, positive and negative elements like those defined for T (see Fig3abcd) in the application to networks.

**Reverse-engineering discrete dynamical systems from disordered observations:**

A cellular automaton (CA) is defined by a rule for computing the new value of each position in a configuration based only on the values of cells in a finite neighborhood surrounding a given position. Commonly a CA evolves on a square grid or lattice of cells updated according to a finite set of local rules which are synchronously applied in parallel. A snapshot in time of the symbols of the cells is called a configuration. A snapshot in space and time (the characteristic CA grid) is called an evolution.

A local and a global function f and $\lambda$ can therefore define a cellular automaton. Let S be a finite set of symbols of a cellular automaton (CA). A finite configuration is a configuration with a finite number of symbols, which differs from a distinguished state b (the grid background) denoted by $0^{\sim}b0^{\sim}$ where b is a sequence of symbols in S (if binary then S = {0, 1}). A stack of configurations in which each configuration is obtained from the preceding one by applying the updating rule is called an evolution. Formally, Let f : $S^Z \rightarrow S^Z$ where Z is the set of positive integers and n, i $\in$ N then $f(r_t) = \lambda(x_{i-r} \ldots x_i \ldots x_{i+r})$, where f is a configuration of the CA and $r_t$ a row with t $\in$ N and $r_0$ the initial configuration (or initial condition). The function f is also called the global rule of the CA, with $\lambda : S^n \rightarrow S$ the local rule determining the values of each cell and r the neighborhood range or radius of the cellular automaton, that is, the number of cells taken into consideration to the left and right of a central cell $x_i$ in the rule that determines the value of the next cell x.

All cells update their states synchronously. Cells at the extreme end of a row must be connected to cells at the extreme right of a row in order for f to be considered well defined. The function $\lambda$ indicates the local state dependency of the cellular automata and f updates every row. Depicted (Extended Data Figures 7-13) is the Elementary Cellular Automaton (ECA) rule 254 (in Wolfram's enumeration[12]) that generates a typical 1-dimensional cone from the simplest initial condition (black cell) running downwards over time for 20 steps. ECA are CA that consider only the closest neighbours to the right and left and itself, thus 3 cells, each with a binary choice for $\lambda$. Every ECA such as rule 254 can thus be seen as a $2^3 = 8$-bit computer program represented by its rule icon representing its function f (Fig1a P(t)) or a function determining its local and global dynamics (Fig. 2 D(t)). Any perturbation of the simple evolution of the rule leads to an increase of its complexity because a rule with a longer description than rule 254 would be needed to incorporate the random perturbation introduced (blue rows). Thus every row in rule 254 is information negative, except for the random rows whose deletion would bring the rule to its simplest description (rule 254). Unlike the rest of the dynamic system, the last step in the evolution of a dynamic system is information neutral because it does not add or remove any complexity, so removal of neutral elements reverses the system's unfolding evolution to its original cause (the black cell) and the rule can be derived by reversing the sequence of the neutral elements at every step, effectively peeling back the dynamic system from a single instant of a sequence of observations (in optimal conditions, e.g. no noise and full accuracy, and good enough approximations of algorithmic-information content).

When clustering consecutive rows of the evolution of all Elementary Cellular Automata (256 rules) we found that the later the perturbations in time, the more neutral, thus conforming to the theoretical expectation (Fig. 4 main text). When taking a sample of representative ECA, this was also clearly the case (Fig. 4 main text). We proceeded to reverse engineer the rule of an ECA by:

1) Producing the space-time diagram O(n) of an ECA from time 0 (initial condition) to time n.
2) Scrambling the observations from O(n) (worse case of an observation, to lose track of their order)
3) Sorting the scrambled observations to maximize algorithmic probability and thus find the most likely generating mechanism (with lowest algorithmic complexity).
4) From 2 and 3 estimating the algorithmic-information content of every (hypothesized) step.
5) Comparing among them and sorting from lowest contribution to highest.
6) Finding the initial condition and generating rule by reversing the order of the sequence of neutral elements from O(n).

Finding the lowest complexity configuration of disordered observations we show how we found the correct times, thus generating a most powerful method to reverse engineer and find design principles and the generating mechanism of evolving systems. Running the sequence forward one can also make predictions about the phase space configuration of the dynamic evolving system. Fig. 3 shows that the predicted point in the phase space does not diverge from the actual position of the system in phase space, thus providing good estimations of the evolution of the system both backward and forward.

In this paper, we choose to work at the level of T (see Fig. 3 main text) for the same convenient simplifying reasons followed by other network-based approaches, but unlike other possible approaches, the theory and methods hold in general for non-linear dynamical systems and not only for static or evolving networks. When working on T only, we assume that lossless descriptions are of the observations (e.g. only T) and not of full descriptions of T and D or even P (the true generating mechanism, e.g. a computer program P) that is the unknown. To date, there have been no other alternatives to applying non-linear interventions to complex systems in the phase space other than to actually calculate the dynamical properties of a system, often assumed with little knowledge or else assumed to be linear and in fixed states, requiring computationally intractable simulations. This new calculus, however, requires much less information to make educated causal interventions that prove to be extremely useful and powerful.

**Entropy-deceiving graphs**

We introduced a method[6 (main text)] for building a family of recursive graphs of which one is denoted by 'ZK' with the property of being recursively constructed and thus of low algorithmic (Kolmogorov-Chaitin-Solomonoff) complexity (hence causal) but that to an uninformed observer would appear statistically random and thus as having maximal Entropy. These graphs were proven to have maximal Entropy for some lossless descriptions but minimal Entropy for other lossless descriptions of exactly the same object, thereby demonstrating how Entropy fails at unequivocally and unambiguously characterizing a graph independent of a particular feature of interest reflected in the choice of natural probability distributions. A *natural probability distribution* of an object is given by the uniform distribution suggested by the object dimension and its alphabet size. For example, if a graph G is losslessly (with no loss of information) described by its adjacency matrix M, then in the face of no other information, the natural distribution is the probability space of all matrices of dimensions |M| and binary alphabet. If, however, G is losslessly described by its degree sequence S, with no other information provided about G, the natural distribution is given by the probability space of all sequences of length |S| and alphabet size |{S}|, where {S} denotes the number of n-ary different symbols in S. The natural distribution is thus the less informative state of an observer with no knowledge of the source or nature of the object (e.g. its recursive character). We denote by 'ZK' the graph (unequivocally) constructed as follows:

1. Let 1 → 2 be a starting graph G connecting a node with label 1 to a node with label 2. If a node with label n has degree n, we call it a *core node*, otherwise, we call it a supportive node.
2. Iteratively add a node n + 1 to G such that the number of core nodes in G is maximized.
3. The resulting graph is typified by the one in Fig3c in the main text.

Clearly, supporting nodes are always the latest to be added at each iteration. Perturbing elements of the network other than the last elements will break the generating program and thus these elements will move the network towards randomness, whereas removing the latest nodes has little to no impact because it only moves the network back in time, the originating program remaining the same and only needing to run again to reach the same state as before. Thus by inspecting elements that do not contribute or make the network slightly simpler, one can reverse the network in time, thereby revealing its generating mechanism (See subsection titled Algorithmic Causal Reconstruction of Dynamic Systems).

We have shown that Entropy is highly observer dependent[6 (main text)], even in the face of full accuracy and access to lossless object descriptions. For these specific complexity-deceiving graphs Entropy produces disparate values when the same object is described in different ways (thus with different underlying probability distributions), even when the descriptions reconstruct exactly the same, and only the same, object. This drawback of Shannon Entropy, ultimately related to its dependence on distribution, is all the more serious because it is often overlooked for objects other than strings, such as graphs. For an object such as a graph, we have shown that changing the descriptions may not only change the values but that divergent, contradictory values are produced. This means that one not only needs to choose a description of interest to apply a definition of Entropy-- such as the adjacency matrix of a network (or its incidence or Laplacian) or its degree sequence--but that as soon as the choice is made, Entropy becomes a trivial counting function of the feature-- and only the feature--of interest. In the case of, for example, the adjacency matrix of a network (or any related matrix associated with the graph, such as the incidence of Laplacian matrices), Entropy becomes a function of edge density, while for degree sequence, Entropy becomes a function of sequence normality. Entropy can thus trivially be replaced by such functions without any loss, but it cannot be used to profile the object (randomness, or information content) in any way, independent of an arbitrary feature of interest. The measures introduced here are robust measures of (graph) complexity independent of object description based upon the mathematical theory of randomness and algorithmic probability (that includes statistical randomness), which are sensitive enough to deal with causality and provide the framework for a causal interventional calculus.

## SI SECTION 2 REFERENCES

## Section 3: Evaluation and validation of the causal calculus using transcriptional data and genetic regulatory networks.

**E-Coli Transcription Factor Network Ontology Enrichment Analysis**

We estimated the information node values of a highly curated E. coli transcriptional network (only experimentally validated connections) from the RegulonDB (http://www.ccg.unam.mx/en/projects/collado/regulondb). Info values were clustered into 6 clusters by using partitioning around K-medoids and optimum average silhouette width. Gene clusters were tested for enrichment of biological functions according to Gene Ontology, KEGG and EcoCyc databases, using the topGO "weight01" algorithm for GO or hypergeometric enrichment test for KEGG and EcoCyc. BDM values did not correlate with degree distribution, compression or Shannon entropy. The numerical results suggest that more positive information genes in E-Coli are related to homeostatic processes, while more negative info genes are related to processes of specialization, which is in agreement with the idea that cellular development is an unfolding process in which core functions are algorithmically developed first, then more specialized functions, enabling training-free and parameter-free gene profiling and targeting. Extended Figures 8-16 show that other measures fell short at producing statistically significant groups for a gene ontology analysis, and also provide details of the clusters found and the elements comprising them.

**Information spectral and enrichment analysis of Th17 differentiation**

We applied our method to a dataset on differentiation of T-helper 17 (Th17) cells[1]. Th17 cells are one of the major subsets of T-helper cells, which in addition to Th17 comprise several sub-types such as Th1, Th2 and Treg cells. These subsets all differentiate from a common naïve CD4+ T cell precursor cell type based on environmental signals and are classified by certain lineage-defining markers. Th17 cells are necessary to protect the host from fungal infections, but at the same time are involved in the pathogenesis of several autoimmune diseases, hence the processes driving Th17 differentiation are of great interest to the scientific community[3]. From the gene ontology analysis taking the experimentally known genes involved in the process of differentiation from T naïve to Th17 (Fig. 4e), it is shown that precisely these genes are distributed non-uniformly and in different ways along the 3 time points, suggesting that the algorithmic perturbation analysis succeeds at identifying such genes (otherwise, the distributions would have appeared uniform in all cases).

**Information spectral analysis**

The information spectral analysis used a reconstructed regulatory network from functional perturbation and transcriptional data corresponding to the Th17 differentiation. The data was divided into three time windows: 0.5 to 2 hours, 4 to 16 hours, and 20 to 72 hours, here referred to as EarlyNet, IntermediateNet and FinalNet respectively. We were interested in investigating whether genes with strongly negative or positive information values would include genes known to be crucial in Thelper cell differentiation and/or novel putative Th17 regulators, and whether these genes would, according to our predictions, change their information content throughout the Th17 differentiation process. We noted that in general, genes classified as having the most positive or negative information values covered several genes known to be involved in T cell differentiation, such as transcription factors from the IRF or STAT families (see Extended Data Figure 5). The genes assigned to the Th17 regulating modules[2] were present along the spread of information values, with some enrichment at extreme positive values. However, not all genes with extreme information values were identified in the original study,[35] suggesting that our method may identify additional regulators (Extended Data Figure 5). When analyzing those genes that are present in all 3 networks and determining their evolution over time (Extended Data Figure 5), we noted that genes for chemokines/chemokine receptors were switching from negative values in EarlyNet to positive values in FinalNet. In the gene group switching from positive in EarlyNet via negative in IntermediateNet back to positive in FinalNet, many transcription factors from the STAT family were represented. Extreme (mostly positive) information values were assigned to many members of the IRF family of transcription factors, which comprises well-known regulators of Thelper differentiation (Extended Data Figure 5), including Th17-inhibiting roles for IRF8 which appears at the top of the lists in IntermediateNet and FinalNet (Extended Data Figure 5). Only three genes were assigned negative information values in FinalNet, namely STAT6, TCFEB and TRIM24, suggesting that removing these might enhance the Th17 profile.

## Clustering

The networks were clustered using the k-means algorithm with 5 clusters per network (Extended Data Figure 5). The list of genes that changed from most negative information values in EarlyNet (cluster 5) toward most positive information values in FinalNet (cluster 1) contained several genes involved in T helper cell subset differentiation and function, for example, HIF1a, FOXO1, IKZF4, IL2, IL21, IL2RA, IL6ST. Conversely, the list of genes with the highest information values in EarlyNet overlapping with the lowest information values in FinalNet was more restricted in number and contained some general transcription factors such as RelA and Jun.

We noted that in general, genes classified as having the most negative or positive information values comprised many genes known to be involved in T cell differentiation, such as transcription factors from the IRF or STAT families, chemokine receptors, cytokines and cytokine receptors. This was particularly evident for networks 1 and 3. When analyzing those genes that have negative information values in network 1 and that change towards positive information values in network 3, we found that the common elements in both lists contain several such genes involved in T helper cell subset differentiation and function, for example, HIF1a, FOXO1, IKZF4, IFNg, IL2, IL21, IL2RA, IL6ST, CXCL10, CXCR3, CXCR5. Interestingly, the list of genes with positive information values in network 1 or with negative information values in network 3 was much more restricted in number and did not overlap, yet contained highly interesting genes. In network 1, these were mostly transcription factors, including several IRFs, STATs as well as RUNX1 and SMAD2, all known to be important in T cell differentiation. The few genes with negative information values in network 3 were STAT6, TCFEB and TRIM24 (interestingly these 3 genes, STAT6, TCFEB, TRIM24 are amongst the few centered around 0, i.e. neutral, in network 1), and it is tempting to speculate that over-activation of these might be able to reprogram differentiated Th17 cells to another lineage. Indeed, STAT6 is a well-known factor in IL-4 response and Th2 induction. Notably, in network 2, which may be viewed as a transition state, 3 genes were assigned the most positive information values and all of these belonged to the IRF family of transcription factors, which comprises well-known regulators of Thelper differentiation, including Th17-inhibiting roles for IRF8 which appears in said list.

## Enrichment Analysis

To assess to what extent our informational spectral analysis identifies genes which are relevant to the differentiation process in Th17 cells, we perform an enrichment analysis based on a literature survey. To this end we collected 9 landmark papers in the Th17 literature [2,4–11].

From each paper a list of genes was extracted (manually), in an attempt to select the set of genes, which the text identified as relevant to Th17 differentiation. The script calculates all the intersections between these sets, with genes at a greater number of intersections given a higher weight as being more relevant in the Th17 literature (the list of genes is in Sup. file output_with_kuchroo.txt). The data is represented in a network diagram (Extended Data Figure 6) where a co-occurrence analysis highlighted genes that were commonly identified across several studies.

The enrichment analysis revealed that positive and negative information elements were not distributed equally, thus indicating that information values were not distributed by chance in any of the three time steps, and that these changed over time according to the theoretical and biological expectations. That is, at early stages the naïve cell has two strong sets of genes that act as handles to steer the network towards or away from randomness, with a larger component of negative elements that indicate signals that are either activating the cells or perturbing cells among the stable naïve cells that are key to the original (undifferentiated steady state) program. Then cells are activated and fewer negative genes are present, while there is a distribution skew of the positive patch towards neutral elements that pinpoint the evolving genes from the cell activation for differentiation (high peak in the enrichment analysis). At the final step the cells no longer have negative elements, indicating that the program has reached a steady state and the cells have been fully differentiated, with all remaining elements either positive or closer to neutral.

### CellNet Waddington landscape
CellNet is a network biology-based computational platform that assesses the fidelity of cellular engineering and claims to generate hypotheses for improving cell derivations[45]. We merged networks of the same tissue type into a single larger entity. The result led to a set of networks of networks of the following 16 Homo Sapiens cell types: B-cell, colon, endothelial, esc (embryonic stem cell), fibroblast, heart, hspc (Hematopoietic stem cells), kidney, liver, lung, macrophage, muscleSkel, neuron, ovary, skin and tcell, each with the following vertex count: 12006, 4779, 5098, 16581, 8124, 6584, 21758, 5189, 4743, 1694, 5667, 6616, 10665, 1623, 3687 and 11914, on which we applied the causal calculus and reprogrammability measures (SI Section 1). A Waddington landscape can be derived from the location in the complexity and programmability quadrants according to the theoretical expectation. According to Fig. 4e, more differentiated cells tend to be closer to x = 0, while non-differentiated ones tend to be farther away, because they start from a state of randomness with shallow attractors and are very sensitive to perturbations but can only move in one direction--towards creating functions represented by structures moving away from randomness. And this is exactly what we found when calculating and plotting the CellNet networks from 16 cell lines in Homo Sapiens. The cells from the CellNet networks were organized into about the same shape as in the theoretical sketch (Extended Data Figure 3) describing their thermodynamic-like behaviour and in agreement with the biological stage expectation placing stem cells in order (hspc and esc) closer to randomness and high in reprogrammability, conforming with the theoretical expectation to have the greatest number of possible shallow attractors, with the network only able to move away from randomness, followed by blood-related cells (bcell and tcell) that are known to be highly programmable and adaptable, followed by the bulk of differentiated cells in the first and second quadrants. The distribution of the (re)programmability of cells as represented by networks from CellNet fits the naturally expected (re)programmability (c.f. Supplement Section 1 and Fig. 4e).

## SI SECTION 3 REFERENCES

## Extended Data Figures

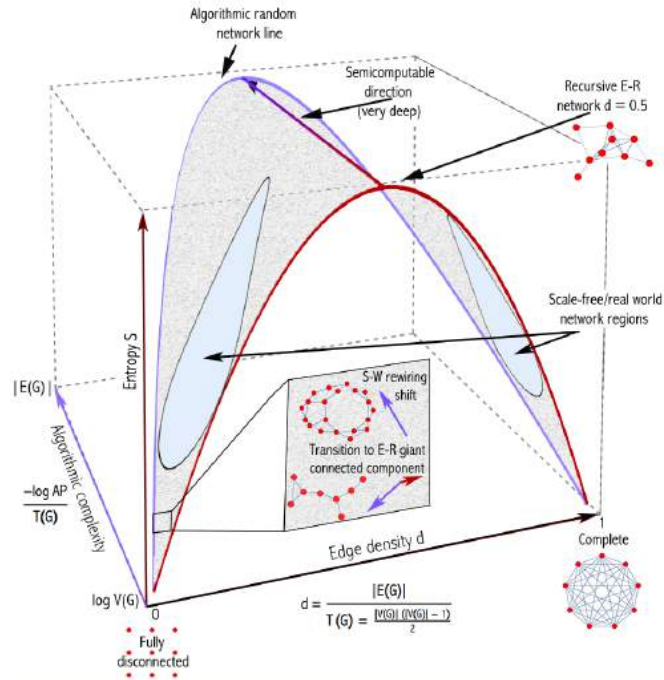

**Extended Figure 1:** Algorithmic complexity (numerically approached by way of Algorithmic Probability) adds an additional dimension (depth), complementary but different from the notion of entropy, when performing network analysis. Unlike statistical mechanical approaches such as Shannon Entropy (for strings or networks), algorithmic complexity improves over Entropy by assigning lower Entropy and thus higher causal content to objects that not only appear statistically simple but also algorithmically simple by virtue of having a short generating mechanism capable of reproducing the causal content of a network. Without such an additional dimension, causal and non-causal networks are collapsed into the same typical Bernoulli distribution. Indeed, a random-looking network with maximal Shannon entropy can be recursively generated by a short algorithm that Entropy would misclassify as random. This additional dimension that we introduce in the study of dynamic systems, in particular networks, together with methods and tools, is thus key to better tackling the problem of revealing first principles and discovering causal mechanisms in dynamic evolving systems. The new dimension can account for all types of structures and properties and is sensitive in both directions, where computable or statistical measures would not be. Indeed, an Erdös-Rényi graph, for example, can be recursive or not, with recursivity meaning that it is actually pseudo-random and only has the properties of a random graph but is not algorithmic-random. This distinction is key in science, where evolving systems may be random-looking but are governed by rules that are otherwise concealed by apparent noise.

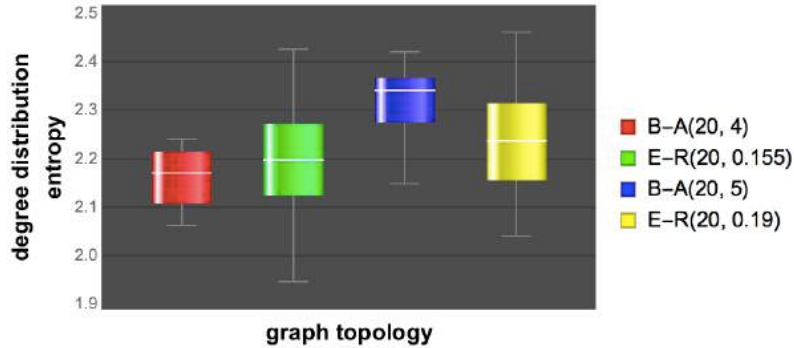

**Extended Figure 2:** Entropy can easily be fooled. Here is a preferential attachment algorithm (B-A) creating networks of growing density (edge number per node) showing Entropy when calculated on adjacency matrices by only capturing graph density, assigning dense B-A graphs higher entropy than Erdös-Rényi (E-R) graphs. This result was reproduced in 30 replicates using 20 node graphs and 20 replicates/graphs and the experiment was repeated approximately 10 times[1]. The main Fig2c shows another graph created recursively (and thus of low algorithmic complexity) that suggests divergent values of Entropy for the same object but with different descriptions, suggesting different probability distributions. A different, more robust approach to characterizing networks and systems is thus needed to be able to tell these cases apart, moving into the algorithmic mechanics/calculus introduced here and thus improving over traditional techniques that draw heavily upon statistical mechanics.

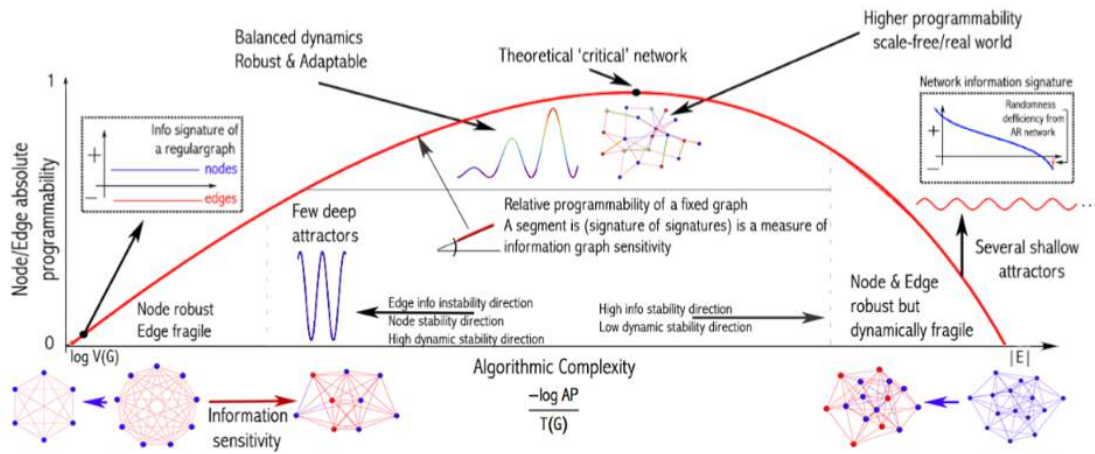

**Extended Figure 3.** A thermodynamic-like effect based on (re)programmability, a measure of sophistication: Moving random networks by edge removal is significantly more difficult than moving simple networks towards randomness. For random graphs, there are only a few elements, if any, that can be used to move it slowly towards simplicity. In contrast, a greater number of elements can move a simple network faster towards randomness. This relationship, described by the reprogrammability rate Δ(G) < Δ(G') (see Sup Mat) for G simple and G' random graphs of the same size (vertex count), induces a thermodynamic-like asymmetry based on algorithmic probability and reprogrammability. A MAR graph, which is of the highest algorithmic randomness, has Δ(MAR) = log n for all its elements after n element removals, and thus cannot be easily moved towards greater randomness. This reprogrammability landscape is thus also expected to be related to the dynamical space (epigenetic) landscape with controlled effects in the phase space according to the complexity and the reprogrammability indices of a system, simple connected graphs having fewer attractors than random graphs of the same size. As we have found and reported in the main text and S.I., moving connected networks towards randomness tends to increase the number of attractors (and therefore make them shallower), providing key insights into the epigenetic Waddington landscape and a tool to move systems and networks hitherto impossible to induce to perform in optimal ways other than by actual simulation. Conversely, moving connected networks away from randomness will tend to reduce the number of attractors (and thereby increase the depth of the remaining ones).

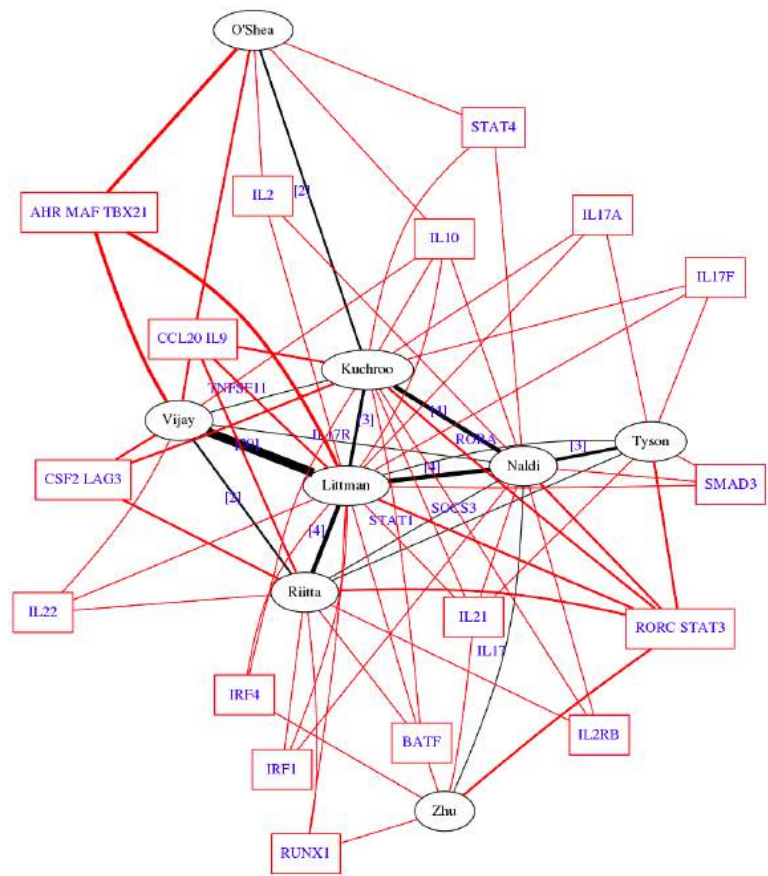

**Extended Figure 4:** Network Venn diagram of genes (square nodes) occurring in the 9 major papers in the literature (black elliptic nodes) covering investigations of Th17 cells [2–10]. These papers cover the majority of genes which have been associated with Th17 cells. Linked genes in the figure are genes found in common between two or more papers. Black lines show the number of genes found in common between two papers (with the thickness denoting the size of the overlap). These genes were used in main Figure 4f,g,h in the gene enrichment analysis of the Th17 differentiation network.

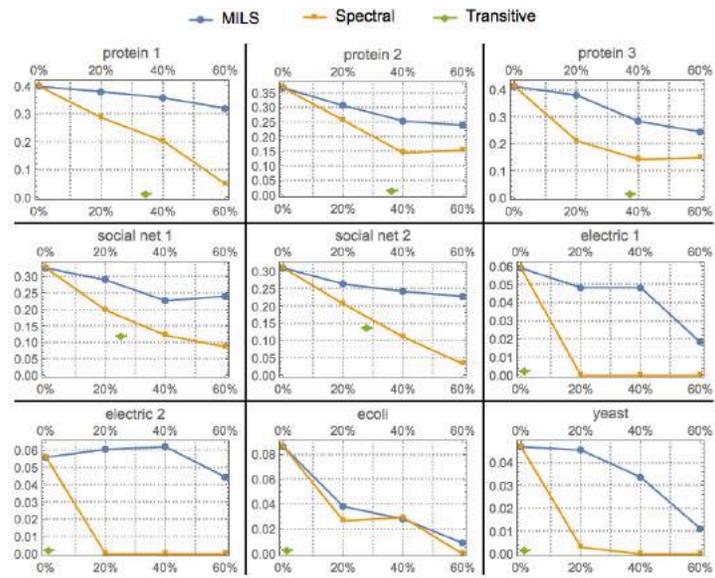

**Extended Figure 5:** Evaluating MILS using 9 benchmark networks common in the literature [11] as regards its ability, compared to two state-of-the-art network dimensionality reduction methods, to preserve the clustering coefficient of the original networks while removing up to 60% of all the network edges [12]. Similarly, MILS preserved edge betweenness, degree distribution (see Main Figure 1 l-p) and information signatures (by design) better than other methods such as random edge/node deletion and lowest degree node deletion. This is to be expected because all these properties of a network are part of its description. MILS thus minimizes the loss of information by maximizing the preservation of all the properties of the original networks.

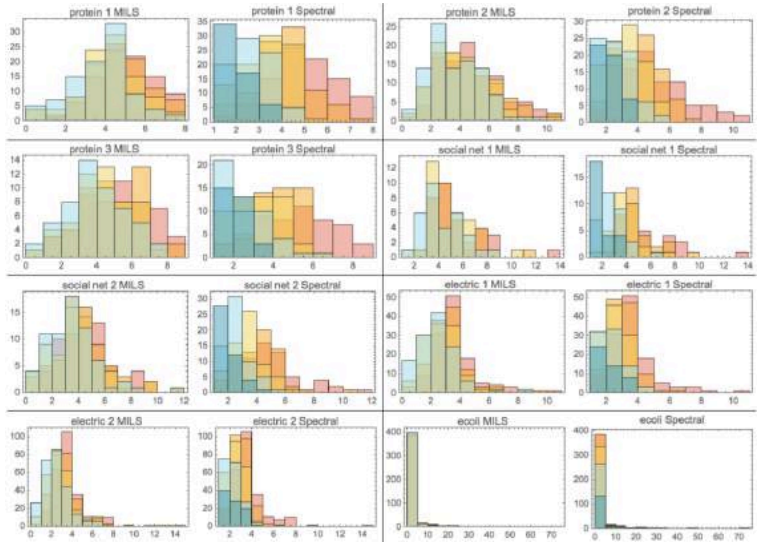

**Extended Figure 6.** Histograms showing the preservation of degree distributions by MILS against a benchmark dimensionality-reduction algorithm based on graph spectra that maximizes the preservation of the graph eigenvalues when removing 20% of the edges (blue), 40% (yellow), 60% (orange) and 80% (pink). The colour green represents the overlapping of areas for each graph and each method. The graphs used are a set of benchmarking graphs in the literature[11].

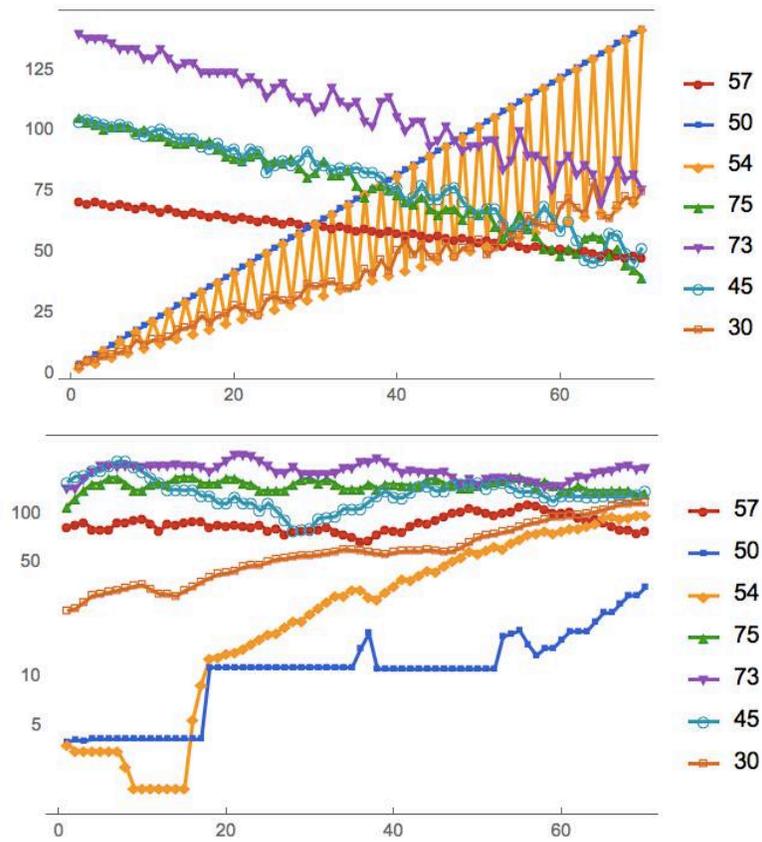

**Extended Figure 7.** Qualitative reconstruction by representing each row in a CA as a binary vector, which produces a 2n+1 dimensional phase space, where n is the CA runtime for a sample of representative ECAs. The hamming distance between the binary vectors is used to calculate the behaviour of the moving particle indicating the state of the ECA (top plot). Applying the same procedure to the hypothesized generating mechanism, as identified from our causal calculus, we find that the moving average (bottom plot) of the predicted particle qualitatively moves in a similar fashion (e.g. increasing v. decreasing/constant) as the original ECA, and the order among the lines corresponds to the original one.

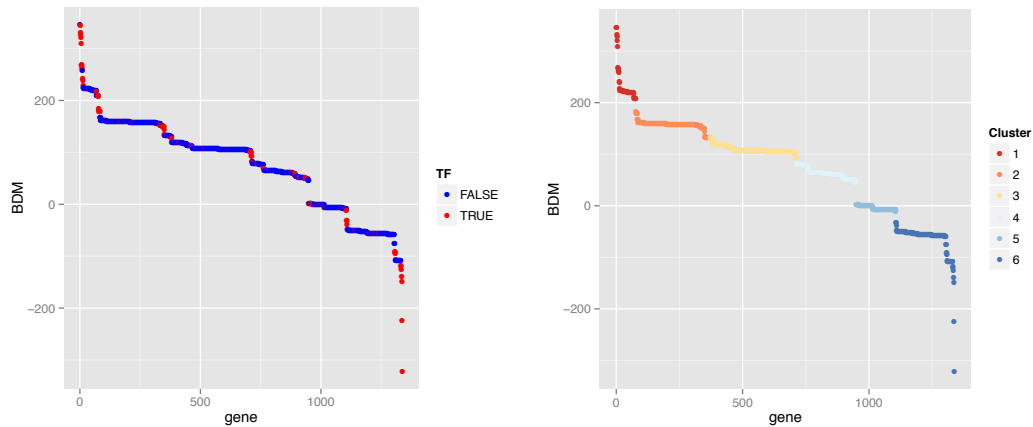

**Extended Figure 8.** Six clusters were selected using partitioning around medoid clustering. The number of clusters was estimated by optimum average silhouette width.

|  | GO.ID | Term | Pval |
|---|---|---|---|
| **Cluster 1** | GO:0006094 | gluconeogenesis | 1.60E-06 |
|  | GO:0006096 | glycolysis | 0.00036 |
|  | GO:0008615 | pyridoxine biosynthetic process | 0.0124 |
|  | GO:0009255 | Entner-Doudoroff pathway | 0.0124 |
|  | GO:0042330 | taxis | 0.02035 |
|  | GO:0016052 | carbohydrate catabolic process | 0.02911 |
| **2** | - | - | - |
| **3** | - | - | - |
| **4** | - | - | - |
| **5** | - | - | - |
| **III** | GO:0006793 | phosphorus metabolic process | 2.10E-08 |
|  | GO:0009252 | peptidoglycan biosynthetic process | 2.90E-07 |
|  | GO:0006777 | Mo-molybdopterin cofactor biosynthetic process | 1.20E-05 |
|  | GO:0009086 | methionine biosynthetic process | 0.0027 |
|  | GO:0009242 | colanic acid biosynthetic process | 0.0124 |
|  | GO:0006164 | purine nucleotide biosynthetic process | 0.0196 |
|  | GO:0009228 | thiamine biosynthetic process | 0.0254 |
|  | GO:0009243 | O antigen biosynthetic process | 0.0254 |

**Extended Figure 9.** Gene Ontology **GO** database (Biological Process category): over-represented categories tested with TopGO weight01 method (Fisher p<0.05)

| | KEGG ID | Term | Pval |
|---|---|---|---|
| **Cluster 1** | 00010 | Glycolysis / Gluconeogenesis | 1.76E-08 |
| | 00051 | Fructose and mannose metabolism | 7.13E-06 |
| | 02030 | Bacterial chemotaxis | 6.32E-05 |
| | 02020 | Two-component system | 7.55E-04 |
| | 00620 | Pyruvate metabolism | 4.08E-03 |
| | 00030 | Pentose phosphate pathway | 5.14E-03 |
| | 02060 | Phosphotransferase system (PTS) | 5.45E-03 |
| | 00680 | Methane metabolism | 6.70E-03 |
| | 01110 | Biosynthesis of secondary metabolites | 9.59E-03 |
| | 01120 | Microbial metabolism in diverse environments | 1.44E-02 |
| **2** | - | - | - |
| **3** | | | |
| **4** | - | - | - |
| **5** | - | - | - |
| **6** | 00550 | Peptidoglycan biosynthesis | 1.01E-07 |
| | 01100 | Metabolic pathways | 6.74E-04 |
| | 04122 | Sulfur relay system | 4.11E-03 |
| | 00621 | Dioxin degradation | 9.20E-03 |
| | 00622 | Xylene degradation | 9.20E-03 |
| | 00360 | Phenylalanine metabolism | 1.48E-02 |
| | 00300 | Lysine biosynthesis | 2.48E-02 |
| | 00230 | Purine metabolism | 3.50E-02 |
| | 00670 | One carbon pool by folate | 3.73E-02 |

**Extended Figure 10.** Over-represented **KEGG** pathways database (p<0.05)

| | EcoCyc pathway | Term |
|---|---|---|
| **Cluster 1** | superpathway of glycolysis and Entner-Doudoroff | 5.37E-07 |
| | Sugar Alcohols Degradation | 4.82E-06 |
| | superpathway of hexitol degradation (bacteria) | 1.91E-05 |
| | glycolysis I (from glucose-6P) | 1.91E-05 |
| | glycolysis II (from fructose-6P) | 1.91E-05 |
| | gluconeogenesis I | 2.56E-04 |
| | Gluconeogenesis | 2.56E-04 |
| | Sugar Derivatives Degradation | 0.003115401 |
| | Secondary Metabolites Degradation | 0.003131693 |
| | superpathway of glycolysis, pyruvate dehydrogenase, TCA, and glyoxylate bypass | 0.004830985 |
| | TCA cycle | 0.004830985 |
| | Glycolysis | 0.005196795 |
| | Generation of Precursor Metabolites and Energy | 0.005701038 |
| | sedoheptulose bisphosphate bypass | 0.037381258 |
| | Entner-Duodoroff Pathways | 0.037381258 |
| | Entner-Duodoroff pathway I | 0.037381258 |
| | CpxAR Two-Component Signal Transduction System | 0.037381258 |
| | Signal transduction pathways | 0.045972995 |
| **2** | - | - |
| **3** | - | - |
| **4** | - | - |
| **5** | - | - |
| **Cluster 6** | methylphosphonate degradation I | 9.40E-06 |
| | Phosphorus Compounds Metabolism | 9.40E-06 |
| | Methylphosphonate Degradation | 9.40E-06 |
| | Pyrimidine Nucleobases Degradation | 0.003167986 |
| | Uracil Degradation | 0.003167986 |
| | uracil degradation III | 0.003167986 |
| | peptidoglycan biosynthesis (meso-diaminopimelate containing) | 0.003167986 |
| | Peptidoglycan Biosynthesis | 0.003167986 |
| | Cell Wall Biosynthesis | 0.003167986 |
| | putrescine degradation II | 0.005063846 |
| | 3-phenylpropionate and 3-(3-hydroxyphenyl)propionate degradation | 0.018877832 |
| | proline to cytochrome bo oxidase electron transfer | 0.019695489 |
| | UDP-N-acetylmuramoyl-pentapeptide biosynthesis I (meso-DAP-containing) | 0.028546946 |
| | UDP-N-Acetylmuramoyl-Pentapeptide Biosynthesis | 0.028546946 |
| | 2-oxopentenoate degradation | 0.04015748 |
| | Putrescine Degradation | 0.0413727 |
| | Pyrimidine Nucleotides Degradation | 0.06959294 |
| | superpathway of ornithine degradation | 0.075477235 |
| | Purine Nucleotides De Novo Biosynthesis | 0.075477235 |
| | superpathway of purine nucleotides de novo biosynthesis II | 0.075477235 |
| | superpathway of arginine, putrescine, and 4-aminobutyrate degradation | 0.09681385 |
| | L-rhamnose degradation I | 0.09815362 |
| | L-rhamnose Degradation | 0.09815362 |

**Extended Figure 11.** Over-represented **EcoCyc pathways** (FDR<0.05)

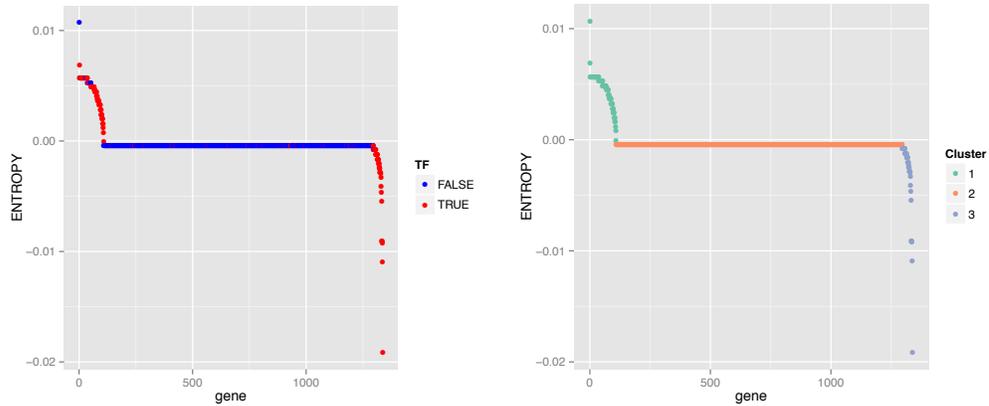

**Extended Figure 12.** Three clusters (above baseline, baseline, below baseline) were identified for Entropy which proved to be less sensitive, clustering most elements over the X axis. Non-baseline nodes are enriched for Transcription Factors.

| | GO.ID | Term | Pval |
|---|---|---|---|
| **Cluster 1** | GO:0006805 | xenobiotic metabolic process | 0.0033 |
| | GO:0009268 | response to pH | 0.0147 |
| | GO:0006355 | **regulation of transcription**, DNA-dependent | 0.0298 |
| **Cluster 2** | GO:0006457 | protein folding | 0.025 |
| **Cluster 3** | GO:0009255 | Entner-Doudoroff pathway | 0.0023 |
| | GO:0009435 | NAD biosynthetic process | 0.0108 |

**Extended Figure 13.** Gene Ontology (Biological Process): over-represented categories tested with TopGO weight01 method (Fisher p<0.05) using Shannon Entropy. No significant groups were found after GO enrichment analysis.

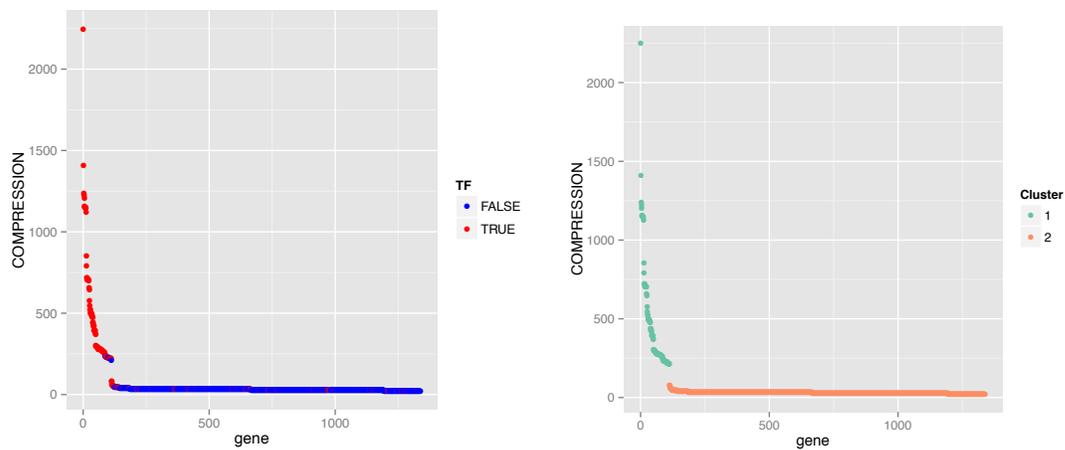

**Extended Figure 14.** Two clusters identified using Compress (above baseline, baseline). Above-baseline nodes are enriched for Transcription Factors. No significant groups were found after GO enrichment analysis.

| | GO.ID | Term | Pval |
|---|---|---|---|
| **Cluster 1** | GO:0006805 | xenobiotic metabolic process | 0.003 |
| | GO:0009255 | Entner-Doudoroff pathway | 0.014 |
| | GO:0006355 | **regulation of transcription**, DNA-dependent | 0.029 |
| **Cluster 2** | - | - | |

**Extended Figure 15.** Gene Ontology (**Biological Process**): Over-represented categories tested with TopGO weight01 method (Fisher $p<0.05$) using lossless compression (Compress algorithm).

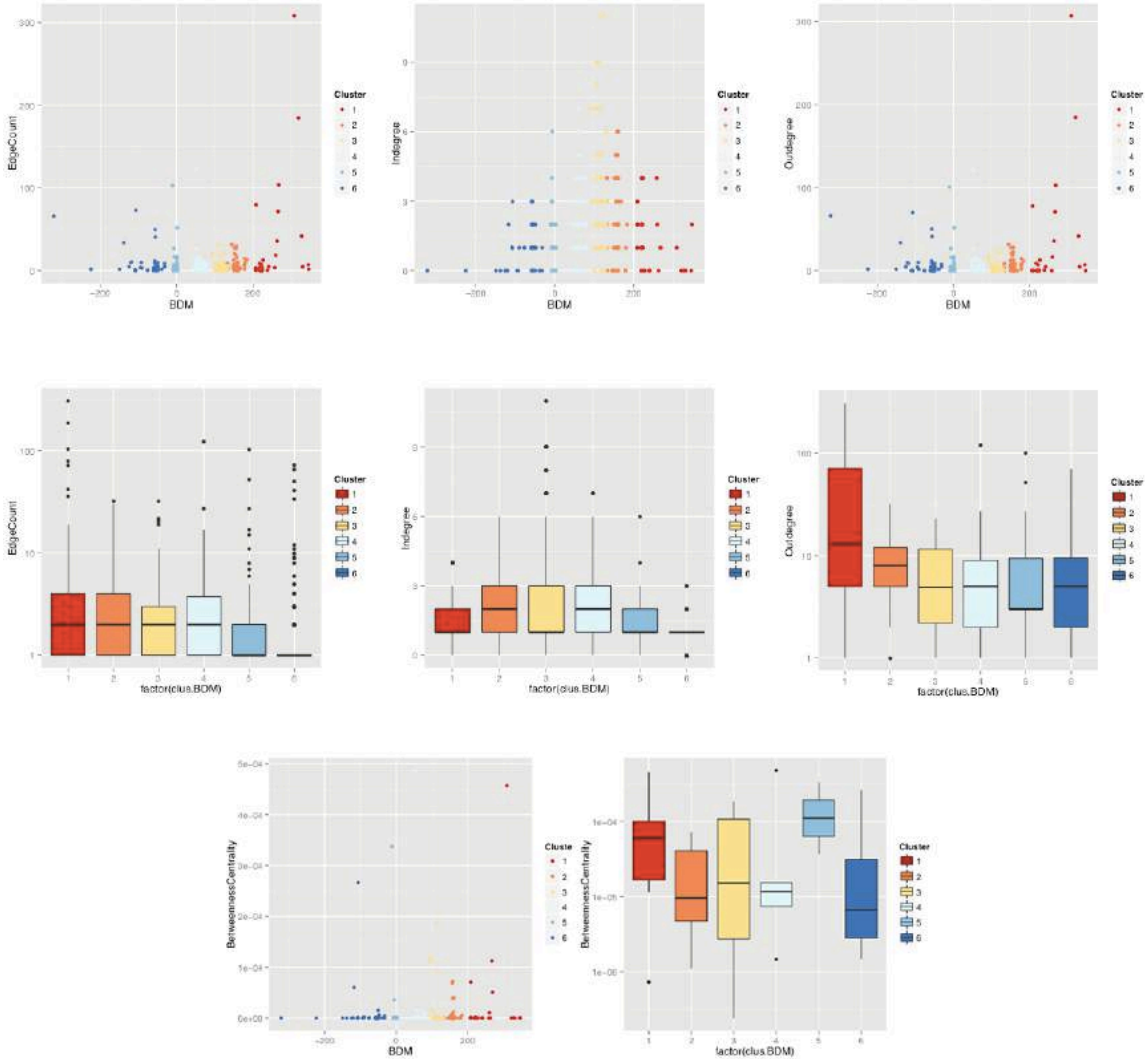

**Extended Figure 16.** Unlike graph-theoretic measures that can be described as single or composed functions of other graph-theoretic measures, BDM was not found to correlate with any of these measures, just as it did not correlate with lossless compression and Shannon entropy. Control Experiments: All attempts to produce statistically significant clusters from graph-theoretic measures, lossless compression and Shannon entropy failed when tested against the same Gene Ontology databases.

## EXTENDED FIGURES REFERENCES